\documentclass{article}

\usepackage{natbib}
\usepackage{amsmath} 
\usepackage{arxiv}
\usepackage[utf8]{inputenc} 
\usepackage[T1]{fontenc}    
\usepackage{hyperref}       
\usepackage{url}            
\usepackage{booktabs}       
\usepackage{amsfonts}       
\usepackage{nicefrac}       
\usepackage{microtype}      
\usepackage{lipsum}
\usepackage{graphicx}
\usepackage{textcomp}
\usepackage{color}
\graphicspath{ {./images/} }
\usepackage{tabularx}
\usepackage{epstopdf}

\AppendGraphicsExtensions{.tif}

\title{Exploratory functional data analysis of multivariate densities for the identification of agricultural soil contamination by risk elements
}

\author{
 Tomáš Matys Grygar \\
  Institute of Inorganic Chemistry\\
  Czech Academy of Sciences\\
  250 68 Řež, Czech Republic \\
  \texttt{grygar@iic.cas.cz} \\
   \And
 Una Radojičić\thanks{the corresponding author}\\
  Institute of Statistics and Mathematical Methods in Economics\\
  TU Wien\\
  Wiedner Hauptstrasse 8-10, 1040 Vienna, Austria \\
  \texttt{una.radojicic@tuwien.ac.at} \\
  \And
 Ivana Pavlů \\
  Department of Mathematical Analysis and Applications of Mathematics\\ Palacký University Olomouc\\
  771 46 Olomouc, Czech Republic\\
  \texttt{ivana.pavlu@upol.cz} \\
   \And
 Sonja Greven \\
  Chair of Statistics\\ 
  Humboldt-Universität zu Berlin\\
  10178 Berlin, Germany\\
  \texttt{sonja.greven@hu-berlin.de} \\
   \And
 Johanna G. Nešlehová \\
  Department of Mathematics and Statistics\\
  McGill University\\
  Montréal (QC), Canada H3A 0B9\\ 
  \texttt{johanna.neslehova@mcgill.ca} \\
   \And
 Štěpánka Tůmová \\
  Faculty of Environment\\
  J. E. Purkyně University in Ústí nad Labem\\
  400 96 Ústí nad Labem, Czech Republic\\
  \texttt{stepatumova@seznam.cz} \\
    \And
 Karel Hron \\
  Department of Mathematical Analysis and Applications of Mathematics\\ Palacký University Olomouc\\
  771 46 Olomouc, Czech Republic\\\texttt{karel.hron@upol.cz} \\
 }

\begin{document}
\maketitle
\begin{abstract} 
Geochemical mapping of risk element concentrations in soils is performed in many countries around the world. It results in numerous large datasets of high analytical quality, which can be used to identify soils that violate individual legislative limits for safe food production. However, there is a lack of advanced data mining tools that would be suitable for sensitive exploratory data analysis of big data while respecting the natural variability of soil composition. To distinguish anthropogenic contamination from natural variation, the analysis of the entire data distributions for smaller sub-areas is key. In this article, we propose a new data mining method for geochemical mapping data based on functional data analysis of probability density functions in the framework of Bayes spaces after post-stratification of a big dataset to smaller districts. The tools we propose allow us to analyse the entire distribution, going well beyond a superficial detection of extreme concentration anomalies. We illustrate the proposed methodology on a dataset gathered according to the Czech national legislation (1990--2009), whose information content has not yet been fully exploited. Taking into account specific properties of probability density functions and recent results for orthogonal decomposition of multivariate densities enabled us to reveal real contamination patterns that were so far only suspected in Czech agricultural soils. We process the above Czech soil composition dataset by first compartmentalising it into spatial units, in particular the districts, and by subsequently clustering these districts according to diagnostic features of their uni- and multivariate distributions at high concentration ends. This allows us to focus on compartments showing known features of contamination. Comparison between compartments, notably neighbouring districts with similar natural factors controlling soil variability, is key to the reliable distinction of diffuse contamination. In this work, we used soil contamination by Cu-bearing pesticides as an example for empirical testing of the proposed data mining approach. In general, there are no natural and justifiable thresholds of risk element concentrations that would be valid for geographical areas with too much natural heterogeneity. Therefore, national (or larger) soil geochemistry datasets cannot be processed as a whole. Empirical knowledge and careful tailoring of statistical tools for the characteristic types of soil contamination are essential for unequivocal identification of the anthropogenic component in real datasets.
\end{abstract}


\section{Introduction}
Detection of anomalies in geochemical data is a traditional task in mineralisation surveys \citep{SINCLAIR1991, Mckinley2016,Borojerdnia2020} and soil and sediment geochemistry studies \citep{Reimann2005, Ander2013, MATYSGRYGAR2016, FabianKarl2017, Lucic2023}. Focus on food production safety and resident health resulted in systematic state-governed mapping of risk element concentrations in soils and the production of highly valuable geochemical datasets in many countries \citep{Ander2013,Vacha2015,Toth2016,Ballabio2018,Reimann2018,Zhang2020}. The existing geochemical datasets cover districts \citep{Galan2008,Zhang2020}, states \citep{Vacha2015,Bednarova2016}, and continents \citep{Toth2016, Reimann2018}. This poses non-negligible challenges for data processing, which must respect specific properties of compositional data \citep{Grunsky2010, Mckinley2016, Zhang2020} and simultaneously produce meaningful results regarding the original purpose of the data gathering \citep{Vacha2015,Toth2016,Greenacre2018,MatysGrygar2023}, which in our case is food production safety. In the case of soil geochemical maps, a common task is to distinguish anthropogenic contamination and natural variability including geogenic anomalies \citep{Ander2013,Amorosi2014,Mckinley2016,MatysGrygar2023}. Agricultural use of anthropogenically polluted areas should be limited if it were to endanger food consumers or residents, but the legislative limits should be applied carefully in natural geogenic anomalies \citep{Amorosi2014}, where risk elements are usually less bioavailable. If the legislative limits for soil contamination were simply set too high so as not to discriminate against owners of land in areas with geogenic anomalies, it could worsen the evaluation of anthropogenic contamination elsewhere. Geogenic anomalies also hinder the identification of diffuse contamination, which can be dangerous in the long term, because weak but persistent worsening of soil composition would be overlooked. This happened recently with Cu \citep{Ballabio2018} and Cd \citep{Reimann2019} in European agricultural soils, for example.\\

In recent work, \cite{MatysGrygar2023} concluded that natural variability in large-scale soil datasets considerably decreases the sensitivity of local anomaly detection that could actually be a consequence of anthropogenic contamination. Geochemical maps can be distorted by variable quartz and organic matter contents \citep{Mckinley2016, Negrel2015}, bedrock composition \citep{Galan2008, Ander2013, Amorosi2014}, and pedogenesis \citep{FabianKarl2017}. If a dataset of soil composition which is too diverse is processed by univariate statistical tools such as the Tukey upper fence (TUF) \citep{tukey1977, Reimann2005, Reimann2018} or Carling upper fence (CUF) \citep{Carling2000,MatysGrygar2023}, the anomaly identification has too low a sensitivity due to too large heterogeneities between the first and the third quartile \citep{MatysGrygar2023}. Many ways to overcome this have been proposed, but most of them focus on processing such big datasets as a whole \citep{Mckinley2016, FabianKarl2017, Ballabio2018, Zhang2020}. Instead,  \cite{MatysGrygar2023} proposed to zoom in on smaller areas, where natural factors, such as parent geology, topography, and climate, are as homogeneous as possible, and to examine the data distribution in the resulting compartments using traditional exploratory data analysis (EDA) tools, in particular empirical cumulative distribution functions (ECDF). The compartmentalisation of big data proposed by Matys Grygar et al. (2023) is similar to the intuitive separation of soil geochemical datasets into subsets (post-stratification) according to the bedrock geology \citep{Galan2008,Zhang2020} and other factors \citep{Ballabio2018, Negrel2015}. \\

When multiple contaminants are being examined, univariate EDA is not sufficient because it ignores possible interactions among elements. In contrast, the Bayes space methodology \citep{Boogaart2014}, allows to process uni- as well as multivariate distributions via their equivalent representation through probability density functions (PDFs). Furthermore, the recent results of \cite{Genest2023} enable an orthogonal decomposition of multivariate PDFs into univariate patterns and interactions among variables. In the present context, this brings additional advantages as multiple contaminants can be studied simultaneously.  
In traditional EDA, soil contamination can be defined as anomalously high element concentrations, separated from a population of non-anomalous (usual) concentrations (main population) by gaps in the data distribution \citep{SINCLAIR1991,Reimann2000,Grunsky2010}. To this end, a population of usual concentrations should be as narrow as possible. This in turn requires the processing of data from areas with rather homogeneous natural factors controlling for soil composition, such as bedrock geology and topography. The practical challenge of such “narrowing” is how to proceed with a mosaic of smaller (more homogeneous) areas without causing an extensive amount of manual work.\\

Most previous attempts to identify soil contamination reduced information from data distribution functions to a threshold (or thresholds) separating usual and anomalous values \citep{Reimann2000,Reimann2018}. Such an approach seems too crude as it fails to take the whole complexity of real soil systems into account. In contrast, in the functional data analysis (FDA) approach we advocate here, the entire probability distribution function can be exploited; a possible information loss is limited to smoothing of the input concentration data via compositional splines \citep{Mach16,Mach21} or kernel density estimation. The latter is preferred here because it is also available in the multivariate case and the resulting density estimates can be used for a subsequent exploratory functional data analysis (EFDA).  \\

\begin{figure}[]
\centering
\includegraphics[width=0.6\textwidth]{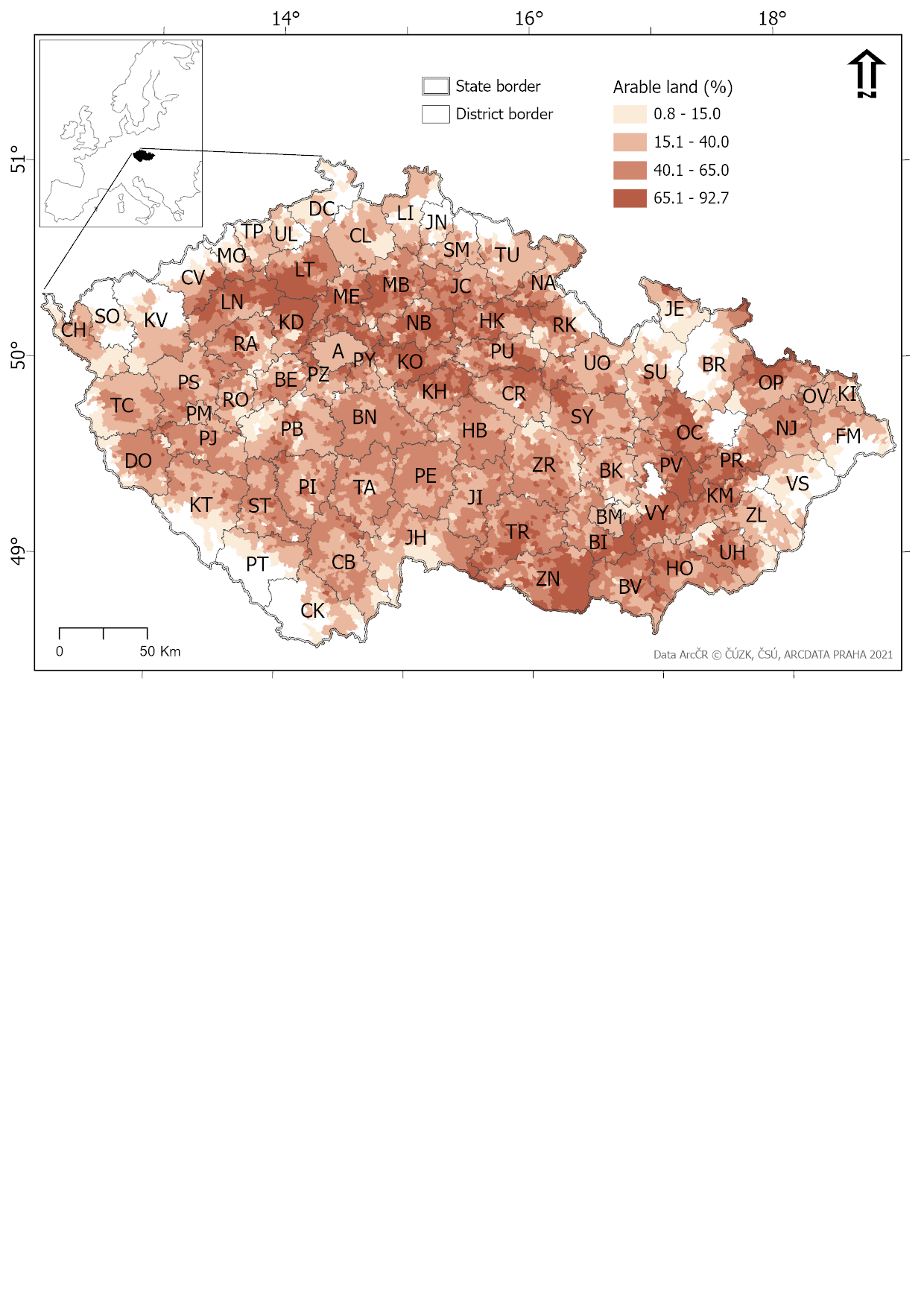}
\vspace{-6cm}\caption{The position and abbreviations of administrative districts in the Czech Republic. The shades of brown show percentages of agricultural fields in the total area of individual cadastral units. }
\label{fig:fig_1}
\end{figure}

The aim of this work is to propose novel, sensitive, and feasible ways to identify anthropogenic contamination of soils by risk elements. The Czech Republic has varied surface geology, topography, and land use. All these factors scatter natural risk element concentrations in agricultural soils. Historical contamination also occurred in the Czech Republic, because it was part of the most industrially developed countries in central Europe around the turn of the 19th and 20th centuries.  Localised contamination is related to ore mining, metallurgy, and heavy industry that caused point contamination of soils; this is relatively well documented in several case studies and overviews of the entire republic \citep{Vacha2015,Bednarova2016,MatysGrygar2023}. The extent of diffuse contamination is less clear but expected due to heavy industry development in the 19th and 20th centuries \citep{Sucharova2014,Vacha2015,Bednarova2016} and agricultural activities with the use of impure fertilisers and pesticides \citep{Skala2020,MatysGrygar2023}. Risk element concentrations in agricultural soils of the Czech Republic have been monitored since the early 1990s \citep{Podleskova1996,Polakova2009} according to the Czech legislation, Act No. 156/1998 Sb. However, the resulting dataset does not include information on natural factors that determine the soil composition, such as bedrock geology, soil texture, and lithogenic element concentrations. This makes the use of this dataset for distinguishing anthropogenic contamination from natural variability challenging \citep{MatysGrygar2023}. The first step made in this paper was to split the entire state database into compartments, in particular into 76 administrative districts (Fig. \ref{fig:fig_1}), in which natural variability is expected to be smaller than in the entire dataset. Risk element patterns in those compartments were examined by advanced tools rooted in the Bayes space methodology and tailored to recognise natural and anthropogenic contamination patterns. This work paves a novel way for the statistical processing of big soil geochemistry datasets through compartmentalisation, clustering similar parts into distinct patterns of contamination, and mode-sensitive identification of contamination by taking into account specific properties of probability density functions from the compartments.

\section{Study area, data, and methods}
\subsection{Study area}

The Czech Republic has an area of 78,700 km$^2$, of which more than a half is used for agricultural purposes. Some of these soils have been contaminated by industrial activities \citep{Vacha2015,MatysGrygar2023}, in particular, historical ore mining and metallurgy and heavy industry in the 19th and 20th centuries. Historical ore mining and metallurgy left clear marks in the soil contamination around the most critical mining cities at the time \citep{Sucharova2014, Bednarova2016, Kylich2022, MatysGrygar2023}. It also concerned some river systems draining ore mining districts and metallurgy centres, such as the Ohře River \citep{Skala2020, Kylich2022}. The major historical metallurgical centres are known, but their actual impact, in particular, diffuse contamination and differentiation between geogenic, metallurgical, and agricultural contamination have not been unequivocally distinguished \citep{Skala2020, MatysGrygar2023}.\\

Administrative districts of the Czech Republic (Figure \ref{fig:fig_1}) have been historically defined to encompass geographically uniform parts of the landscape and thus their geomorphology and geology are relatively consistent. The homogeneity of those compartments is favourable for detailed examination of element distribution functions, similarly as in historical ECDFs proposed for ore prospection \citep{SINCLAIR1991}. In the Czech Republic, there are 76 districts (besides the capital city), which is enough to zoom in from the overall area to compartments, while still having a sufficiently large number of samples per compartment for EFDA, except perhaps for some big cities with a small amount of agricultural soils. Fractions of agricultural soils are unevenly distributed in individual districts (Figure \ref{fig:fig_1}).\\

Preceding case studies in the Czech Republic revealed typical patterns of soil contamination and geogenic anomalies for Cu, Pb, and Zn summarised in Table \ref{tab:tab_1}. All non-ferrous ores mined in the Czech Republic were sulphidic, and polymetallic, with Ag accompanied by Cu (Kutná Hora) or Zn (in all other ore regions) and Pb usually accompanied by Zn \citep{MatysGrygar2023}. Iron ore processing and iron and steel production were linked to emissions of Pb \citep{Sucharova2014} and to a lesser extent Zn, but never Cu due to the contrasting melting points of those metals and the volatility of their oxides. Mixing of contamination sources in districts is not exceptional: the floodplain of the Ohře River has laterally deposited sediments contaminated by half a millennium of mining in the Ore Mountains and hop gardens contaminated by Cu pesticides \citep{Skala2020}. Several other contaminated floodplains used for agriculture are also present in the Czech Republic \citep{Skala2020,MatysGrygar2023}.

\begin{table}[h]
\caption{ Major contamination sources and resulting typical patterns of elevated Cu, Pb, and Zn in soils of the Czech Republic according to preceding studies \citep{Sucharova2014, Bednarova2016, Skala2017, Skala2020, MatysGrygar2023}. Types 1 to 5 are anthropogenic sources; type 6 is a geogenic anomaly and type 7 stands for “uncontaminated” districts.
}
\label{tab:tab_1}
\centering
\begin{tabular}{  |c | m{2.5cm}| 
m{2.5cm} | m{2.5cm} |m{2.5cm} |m{2.5cm} | } 
 \hline
 \textbf{Type} & \textbf{Contamination source} & \textbf{Typical districts} & \textbf{Contaminants} & \textbf{Characteristic element pattern} &  \textbf{Possible overlap}\\
 \hline\hline
 1  & Historical Ag and Pb smelting & 
JI, KH, PB, TA & 
Pb, mostly with Zn, occasionally with Cu & 
High Pb outliers close to smelters plus diffuse contamination & 
Iron/steel metallurgy, geogenic anomalies\\
 \hline
 2  & Pesticides for vineyards, hop gardens, and orchards &
BV, HB, HO, LN, RA, ZN &
Cu never accompanied by Pb and Zn &
Diffuse contamination, high Cu values in Q4 or even in Q3 already &
Cu ore mining and metallurgy (exceptional in the Czech Republic)
 \\
\hline
 3&
Coke, iron, and steel making&
FM, KD, KI&
Pb > Zn, never with Cu&
Diffuse contamination without high outliers &
Ag and Pb smelting \\
\hline
4&
Cities with historical factories, individual industrial sources &
A, BM, OC, OV, PM, PY, PZ, UL&
All three elements&
Frequent elevated Pb, Zn, and/or Cu including high outliers&
Polymetallic ore mining\\
\hline
5&
Polymetallic ore mining
&JE, HB, PB&
All three elements&
Frequent elevated Cu, Pb, and/or Zn
&Large cities with historical factories\\
\hline
6&
Granitic and metamorphic rocks in mountains along the northern state border, also metallogenic zones in Ore Mountains&
JN, KV, LI, MO, TP, TU&
Pb&
Pb elevated in large areas&
Lead glass making, long-range atmospheric deposition\\
\hline
7&
“Almost natural” - minor contamination from individual sources&
CB, CK, CR, JH, PE, PS, TA, VS, ZL
&Pb > Zn > Cu
&Individual outliers (a few \% of samples)&\\
\hline
\end{tabular}
\end{table}

\subsection{Dataset of soil data and geographic information systems}

The Register of Contaminated Areas (RKP) is a database of risk element concentrations in agricultural soils of the Czech Republic, gathered by the state Ministry of Agriculture (MZe) to guarantee food production safety. From RKP, Cu, Pb, and Zn concentrations in soil extracts by cold 2 M HNO$_3$ \citep{Podleskova1996, Zbiral2004, Polakova2009} were used. Although this kind of extraction can be considered historical, as it was later replaced by aqua regia extraction (AR), it covered rather evenly the agricultural lands of the Czech Republic. RKP entries from HNO$_3$ extracts were obtained between 1990 and 2009 and included 49,567 analyses of Cu, Pb, and Zn. Cold 2 M HNO$_3$ extraction recovers ca. 50 \% AR content of Cu, 80 \% AR content of Pb, and 30 \% AR content of Zn \citep{Vacha2013, MatysGrygar2023}. The advantage of the cold diluted HNO$_3$ extraction is higher extractability of atmospheric and fluvial contamination relative to geogenic portions of risk elements \citep{Vacha2013} and generally better relationship with the bioavailable portion of risk elements than pseudo-total or total concentrations \citep{Podleskova1996, Groenenberg2017}. At least one sample per km$^2$ of agricultural soils was obtained from the top 40 cm in ploughed fields, vineyards, hop gardens, and orchards and 30 cm in grasslands. Sampling followed several strategies: the initial one was designed to characterise the entire area of the Czech Republic, and later more detailed analyses were targeted in areas revealed as contaminated. \\

The entire Czech Republic dataset was divided into districts. The district borders, areas of vineyards, hop gardens, and arable land were available from Digital Vector Database of Czech Republic Arc\v CR$^{\mbox{\textregistered}}$ 4.0, \textcopyright\ \v C\'UZK, \v CS\'U, ArcDATA Prague 2021 \url{www.arcdata.cz/cs-cz/produkty/data/arccr}. Geoinformation data processing was performed with ArcGIS Pro, v3.1 software (Esri, USA).

\subsection{Data processing by traditional univariate EDA}
Boxplots and ECDFs are used as conventional EDA tools and constructed using OriginPro software (OriginLab, USA). Quartiles in distributions are denoted conventionally Q1 to Q4. Q2 and Q3 are also highlighted in ECDFs and considered as roughly corresponding to the main population of samples. For some purposes, interquartile ranges (IQR = Q3~--~Q1) are used. The Tukey upper fence (TUF) is defined as 1.5 $\times$ IQR on the linear scale.

\subsection{Statistical processing of PDFs using Bayes spaces}

The relative character of probability density functions (PDFs), which is best conveyed by their common unit integral representation, can be captured by the Bayes space methodology \citep{egozcue06,Boogaart2014}. Being a Hilbert space, a Bayes space provides a convenient algebraic-geometrical structure for scale-invariant functions. Scale invariance of PDFs means that given a common domain and a positive real multiple $c > 0$, two positive functions $f$ and $g$ that are proportional in the sense that $g(x) = c f (x)$ for all $x$ carry essentially the same, relative information. From this perspective, imposing the usual unit-integral constraint of PDFs simply chooses a proper representative in the corresponding equivalence class of scale-invariant functions, as is also known from the concept of proportionality in Bayesian statistics. Consequently, we set up the corresponding Bayes space $B^2(I)$ as a space of such equivalence classes of PDFs. Because we focus on trivariate PDFs for Cu, Pb, and Zn in this study, it will suffice to restrict our presentation to the trivariate case. We will further assume that all trivariate densities have common support of the form $I=I_1 \times I_2 \times I_3 =(a_1,b_1)\times (a_2,b_2)\times(a_3,b_3)$ for some real intervals $(a_1,b_1),\,(a_2,b_2),\,(a_3,b_3)$. The Bayes space  $B^2(I)$ is then defined as the space of all trivariate densities on $I$ whose logarithm is square integrable. The vector space structure of $B^2(I)$ results from introducing operations of perturbation and powering (standing for the usual addition and multiplication operations of functions), defined for $f,\, g \in B^2(I)$ and $\alpha \in \mathbb{R}$ as $(f \oplus g)(x) = f (x) \cdot g(x)$, $(\alpha \odot f )(x) = \alpha f (x)$, respectively \citep{egozcue06}. Note that the result of these operations can be arbitrarily rescaled due to the scale invariance of PDFs. The difference between two elements in the Bayes space can be obtained as the perturbation of one element with the reciprocal of another, i.e. $(f \ominus g)(x) = f (x)\oplus((-1)\odot g(x))$. To endow the $B^2(I)$ with a Hilbert space structure, \cite{egozcue06} defined the inner product $\langle f, g\rangle = 1/(2\eta) \int_I\int_I \ln f(x)/f(y) \ln g(x)/g(y) \mathrm{d}x \mathrm{d}y$, where $\eta=(b_1-a_1)(b_2-a_2)(b_3-a_3)$. The introduced inner product naturally defines a norm and distance on $B^2(I)$ as
$\|f\|=\sqrt{\langle f,f \rangle}$ and $\mathrm{d}(f,g)=\|f-g\|$, respectively.\\

The centred logratio (clr) transformation allows to express a PDF $f \in B^2(I)$ as an equivalent real function in the standard $L^2(I)$ space of square-integrable functions on the cuboid $I$, given as $clr(f) = \ln f - 1/\eta \int_I \ln f(x) \mathrm{d}x$. Although the clr transformation imposes an additional zero integral constraint on the transformed densities, it enables further statistical processing of PDFs using popular FDA methods, which are commonly designed for functions in an $L^2$ space.\\

To distinguish unusual properties of one contaminant from those of two or more contaminants, we will decompose the trivariate densities into their so-called geometric marginals and their interactive parts. Geometric marginals are similar to the usual (arithmetic) marginal densities known from standard probability theory, but marginalise the PDFs on the log-scale so as to be compatible with the clr transformation. The idea is similar to that of the geometric mean, where the relative (ratio) scale of positive data is maintained by computing the arithmetic mean of log-transformed observations and transforming it back using the exponential function. 
\\

Specifically,  in the three-dimensional case, the univariate geometric margins are given by 
\begin{align*}
f_1 & =\exp\left[ \frac{1}{(b_2-a_2)(b_3-a_3)} \int_{I_2 \times I_3} \ln f(x_1,x_2,x_3) \mathrm{d}x_2 \mathrm{d}x_3 \right],\\
f_2 & =\exp\left[ \frac{1}{(b_1-a_1)(b_3-a_3)} \int_{I_1 \times I_3} \ln f(x_1,x_2,x_3) \mathrm{d}x_1 \mathrm{d}x_3\right],\\
f_3 &=\exp\left[ \frac{1}{(b_1-a_1)(b_2-a_2)} \int_{I_2 \times I_1} \ln f(x_1,x_2,x_3) \mathrm{d}x_2 \mathrm{d}x_1 \right]. 
\end{align*}
Consequently, the bivariate geometric margins are defined as $f_{12}=\exp[ 1/(b_3-a_3) \int_{I_3} \ln f(x_1,x_2,x_3) \mathrm{d}x_3 ]$, $f_{23}=\exp[1/(b_1-a_1) \int_{I_1} \ln f(x_1,x_2,x_3) \mathrm{d}x_1]$ and $f_{13}=\exp[ 1/(b_2-a_2) \int_{I_2} \ln f(x_1,x_2,x_3) \mathrm{d}x_2]$. Unlike the usual arithmetic marginals, geometric marginals are affected by the 
dependence structure of the trivariate density \citep{Genest2023}. An important geometric property of geometric marginals is that these are, in fact, orthogonal projections of the original trivariate PDF $f$ onto a suitable lower-dimensional subspace, as shown by \citet{Genest2023}. These authors exploited it to derive the following key decomposition, which is an analogue of the Hoeffding--Sobol identity \citep{hoeffding48,Sobol1993},
\begin{equation}\label{eq:decomposition}
 f=(f_1 \oplus f_2 \oplus f_3) \oplus (f_{12}\ominus f_1\ominus f_2)\oplus (f_{13}\ominus f_1 \ominus f_3)\oplus (f_{23}\ominus f_2\ominus f_3)\oplus (f\ominus f_{12}\ominus f_{13}\ominus f_{23}\oplus f_1\oplus f_2\oplus f_3).
\end{equation}
The first term $f_1 \oplus  f_2 \oplus  f_3$ in the above expression is known as the independence part and equals to the product of the univariate (geometric) marginals; the rest constitutes the so-called interactive part. Note that when the trivariate density corresponds to  stochastically independent variables, the univariate geometric marginals coincide with their arithmetic counterparts, and the interactive part is simply the uniform density (i.e.\ absent from the above decomposition). Accordingly, analysing the usual arithmetic marginals alone neglects possible interactions (either bivariate or trivariate) among variables. In contrast,  geometric marginals could be considered as arithmetic marginals extended by the information about the relationship of a particular element to the others. In the above decomposition, the bivariate interaction between $X_1$ and $X_2$ corresponds to $f_{12}\ominus f_1 \ominus f_2$, i.e., to the bivariate geometric margin $f_{12}$ from which the respective univariate geometric marginals have been filtered out. The bivariate interactions between $X_1$ and $X_3$, resp. $X_2$ and $X_3$ are constructed analogously. Finally, the trivariate interaction results from filtering out all univariate and bivariate geometric marginals from $f$ itself.\\

All elements of the decomposition \eqref{eq:decomposition} given in brackets are mutually orthogonal \citep{Genest2023}. This justifies the use of FDA techniques on each of these terms separately and enables us to effectively identify the source of dependence between the variables. Probably the most important practical implication of the orthogonality of the terms in the decomposition \eqref{eq:decomposition} is the Pythagorean theorem for norms, viz.
\begin{multline*}
 \|f\|^2=\|f_1\|^2 + \|f_2\|^2 + \|f_3\|^2 + \|f_{12}\ominus f_1\ominus f_2\|^2+ \|f_{13}\ominus f_1 \ominus f_3\|^2 + \|f_{23}\ominus f_2\ominus f_3\|^2\\ + \|f\ominus f_{12}\ominus f_{13}\ominus f_{23}\oplus f_1\oplus f_2\oplus f_3\|^2,
\end{multline*}
where for $f \in B^2(I)$, $\|f\|$ denotes its norm in the Bayes space as defined above. \\

In the context of Bayes spaces, the norm of a PDF can be considered as a scalar measurement of information \citep{egozcue18}. Specifically, the PDF of the uniform distribution has zero norm; the more information a density carries (resulting usually in peaky PDFs), the higher the value of its norm. Accordingly, a lower norm of the density indicates the presence of multimodality or heavy tails resulting, e.g., from anthropogenic contamination. The Pythagorean theorem then allows for the decomposition of the overall information, contained in the joint density, into information carried by single variables (through the respective marginals) and their interactions; see also Section 5.4 in \citet{Genest2023}. Moreover, to reflect the above interpretation, the vector of squared norms, viz.
\begin{multline}\label{eq:ic}
   \Bigl(\frac{\|f_1\|^2}{\|f\|^2},\frac{\|f_2\|^2}{\|f\|^2},\frac{\|f_3\|^2}{\|f\|^2}, \frac{\|f_{12}\ominus f_1\ominus f_2\|^2}{\|f\|^2}, \frac{\|f_{13}\ominus f_1 \ominus f_3\|^2}{\|f\|^2}, \frac{\|f_{23}\ominus f_2\ominus f_3\|^2}{\|f\|^2}, \\
     \frac{\|f\ominus f_{12}\ominus f_{13}\ominus f_{23}\oplus f_1\oplus f_2\oplus f_3\|^2}{\|f\|^2}\Bigr)
\end{multline}

is called information composition and will be used under this name in the sequel.\\

One of the most straightforward, yet naive, approaches to the analysis of multivariate density data is to ignore the dependence structure and to focus on the individual analysis of the distribution of the concentration values of the observed elements independently. This amounts to considering the arithmetic marginals of the trivariate density $f$. As the first part of EFDA, we thus focus on the univariate arithmetic marginals, which we analyse using hierarchical clustering and the adaptation of the DDC algorithm for univariate functional data described below. \\

Of course, the dependence structure among variables cannot simply be ignored. Given the existence and complexity of the dependence structure between the concentrations of the three elements we study, it is necessary to identify suitable tools that are easy to grasp by practitioners in a geochemical context. The first source of information on the dependence structure is at hand: the information composition vector in \eqref{eq:ic}. Because the elements of this vector represent relative contributions to the overall information carried by the (squared) norm of $f$, they carry relative rather than absolute information. The very same property is characteristic of compositional data \citep{aitchison86,pawlowsky15,filzmoser18}. For each trivariate PDF, we thus use the seven-component information composition vector in \eqref{eq:ic} for subsequent analysis, in particular in the hierarchical clustering (see Section \ref{sec:multi}). 

\subsection{Deviating Data Cells algorithm}
\label{sec_ddc}
\begin{figure}[]
\centering
\includegraphics[width=0.7\textwidth]{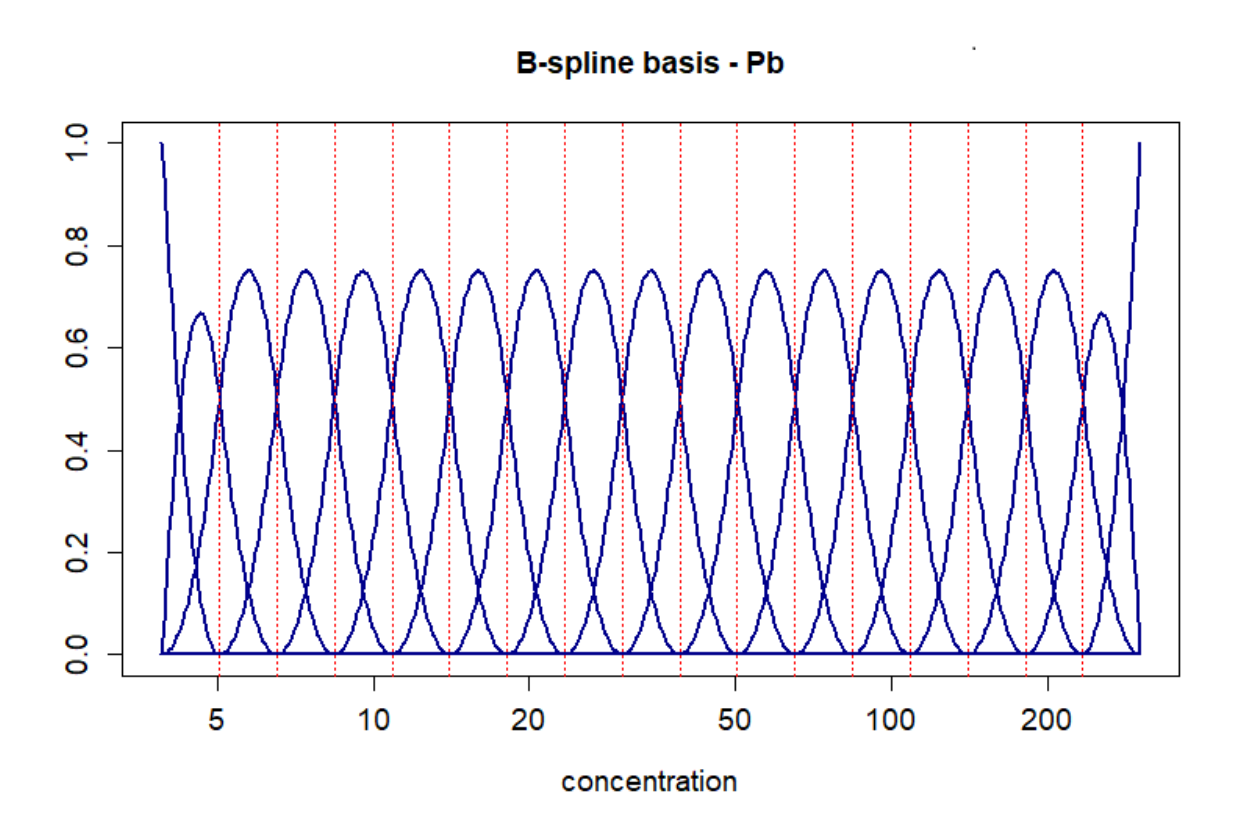}
\caption{One concrete choice of a B-spline basis. For each point on the $x$-axis,  the function values of the basis functions sum up to $1$.}
\label{fig:fig_2}
\end{figure}

The Deviating Data Cells (DDC) algorithm for detecting cellwise outliers in multivariate data was introduced by \citet{rousseeuw18}. It has since proved to be a useful tool for the identification of outlying cells in the data matrix which takes correlations between variables into account. The algorithm can be briefly summarised in the following steps:

\begin{enumerate}
    \item \textit{Standardisation of the variables by means of robust statistics}: Robust estimators of location and scale are used to standardise the observed values within the data matrix. Commonly, the column-wise median is subtracted from the original observed value and the difference is then divided by the column-wise median absolute deviation (MAD).
    \item \textit{Flagging outliers in terms of individual variables}: A cell is marked as an outlier when its absolute value exceeds the cut-off value  $\sqrt{\chi^2_1(p)}$, where  $\chi^2_1(p)$ is the $p$-th quantile of the chi-squared distribution with one degree of freedom.
   \item \textit{Computation of predicted values}: Predicted values for each variable are based on the set of correlated variables, i.e. those with absolute robust correlation higher than 0.5. Each cell is then predicted from the entries of the correlated variables in the same row.  Subsequently, a deshrinking step returns the values to the original scale. In case there are no correlated variables to be used for prediction, the predicted value is set to 0.  Residuals $r_{ij}$, given as the standardised difference between observed and predicted values are then again compared to $\sqrt{\chi^2_1(p)}$ in absolute value -- if exceeded, the cell is marked as an outlier.
    \item \textit{Identification of the rowwise outliers}: The identification of a subject as a row-wise outlier (i.e. outlier at the observation level) is based on a test statistic obtained as a suitable function of the squared residuals $r_{ij}^2$ \citep[equation (12)]{rousseeuw18}. The realisation of this statistic is then compared to the respective criterion $\sqrt{\chi^2_1(p)}$. When exceeded, the row as a whole is then marked as an outlier.
\end{enumerate}
The extension of the DDC algorithm to the case of (univariate) PDFs -- here, in the form of arithmetic marginals -- is quite natural. As commonly used in FDA, we first represent each functional observation as a linear combination of predefined basis functions, and apply the DDC algorithm to the vector of the corresponding weights or coefficients. Using the B-spline basis \citep{deBoor1978} proved to be an efficient tool for nonperiodic functional data, with ZB-splines (compositional splines; \citep{Mach16,Mach21}) being introduced specifically for the case of clr-transformed PDFs. Although the use of ZB-splines is technically more straightforward, we opted for an alternative approach which is preferable here for interpretation purposes. The latter uses the usual B-spline basis in the clr space with additional zero-integral constraints. Note that the values of the basis functions range between 0 and 1 at each individual point in the domain and that the coefficients sum up to 1. The latter can thus be perceived as naturally obtained weights for distinguishing the ‘level’ of anomaly for that point. One concrete choice of the B-spline basis is illustrated in Figure \ref{fig:fig_2}. This way, by applying the DDC algorithm on the matrix of the B-spline coefficients, one can not only locate the parts of densities that are outlying from the common behaviour but also quantify the ‘degree of outlyingness’ at the pointwise level.

\subsection{Clustering of districts}
As discussed above, the grouping of districts according to their similarity is an important part of the analysis. To this end, we employ hierarchical clustering \citep{Nielsen2016}. The procedure begins by considering each individual observation as an individual cluster. Subsequently, it iteratively identifies the two clusters that exhibit the closest proximity to each other and proceeds by merging these two clusters together. The proximity of clusters is, in our case, measured by a suitable distance of the two furthest elements in the two clusters. This is also known as complete-linkage clustering; see, e.g., \cite{Everitt2001}. The procedure is iterated until all the observations are placed in the same cluster. To perform the clustering, we utilised the \texttt{gplot} package in \textsf{R} \citep{qplots}. The output is presented as a dendrogram, which is supplemented here by a heatmap to offer a better visual representation of the clustered data. \\

When clustering the districts based on the squared Pythagorean norm decomposition, the districts are first characterised using the clr transformed seven-dimensional vector of squared norms in \eqref{eq:ic}. We then use these vectors to group the districts into several most similar groups using the complete-linkage clustering algorithm with the Euclidean distance. This way, districts are clustered based on the amount of information carried by the individual components in the decomposition \eqref{eq:decomposition}. When using arithmetic marginals to characterise the distribution of the concentration values in the districts, we apply the complete-linkage clustering algorithm directly to the corresponding arithmetic margins. The latter are then grouped by their similarity based on the matrix of Bayes distances. 

\section{Results}
\subsection{Univariate data analysis by ECDF and high concentrations as possible contamination}

\begin{figure}[t!]
\centering
\includegraphics[width=0.75\textwidth]{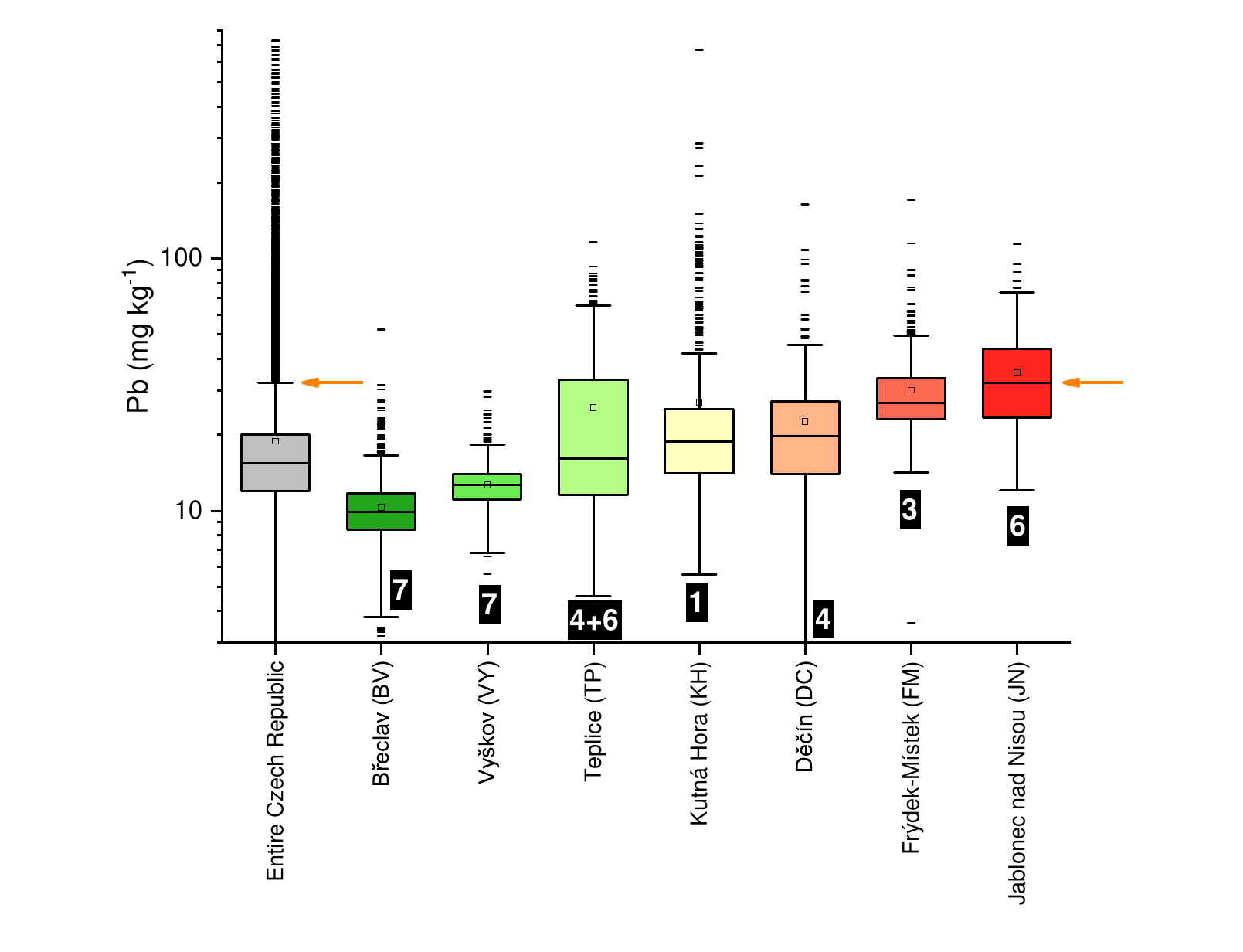}
\caption{Boxplots of Pb concentrations pooled across the entire Czech Republic (left, in grey) along with seven districts covering typical patterns observed in that country. The numbers (white ink in black boxes) refer to contamination patterns according to Table \ref{tab:tab_1}. 
Orange arrows highlight TUF for the entire Czech Republic.}
\label{fig:fig_3}
\end{figure}

\begin{figure}[t!]
\centering
\includegraphics[width=0.9\textwidth]{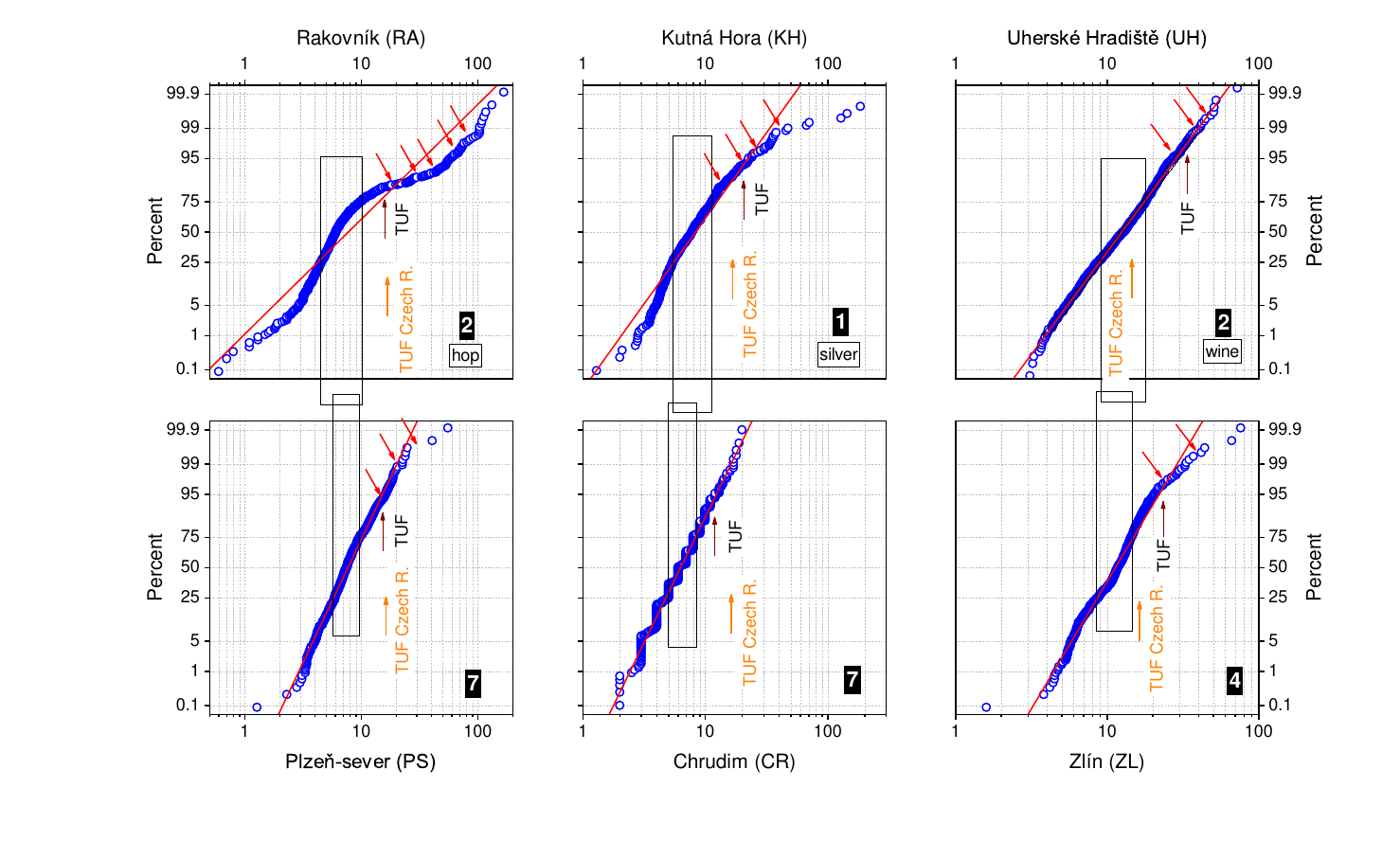}
\caption{ECDFs of Cu (mg kg$^{-1}$) for contaminated (top line) and neighbouring uncontaminated districts (bottom line). Vertical rectangles are boundaries of Q2 and Q3 in individual districts. Numbers (white ink in black boxes) refer to contamination patterns according to Table \ref{tab:tab_1}. TUFs are shown by black arrows for individual data sets and orange arrows for the entire Czech Republic. Red arrows indicate heterogeneities at high concentration levels.}
\label{fig:fig_4}
\end{figure}

The overall variability of risk element concentrations in selected districts and between districts is demonstrated in boxplots (Figure \ref{fig:fig_3}) and ECDF plots (Figures \ref{fig:fig_4} and \ref{fig:fig_5}). The main mass of the concentration distributions, usually corresponding roughly to Q2 and Q3, is quite variable. A comparison of boxplots for selected districts and the entire Czech Republic (Figure \ref{fig:fig_3}) documents the necessity of zooming in (compartmentalisation) from the whole RKP to districts. For example, Teplice (TP) and Jablonec nad Nisou (JN) districts have a high background of Pb, but generally rather low numbers of values exceeding the district-specific TUF. This is consistent with mostly natural causes of the high concentrations: the JN district represents a natural contamination Type 6 according to Table \ref{tab:tab_1} due to a prevailing portion of granitic bedrocks. Application of the entire-Czech Republic TUF (orange arrow in Fig. 3) would thus mark the entire Q3 and Q4 of JN district as anomalous, which is not the case in the context of the district. On the other hand, the entire Czech Republic TUF would be insensitive to identify outliers in the low-background BV and VY districts. IQRs are also district-specific -- in Teplice (TP), the Pb concentration IQR is extremely high (Figure \ref{fig:fig_3}) due to a considerable portion of anomalous soils from the Ore Mountains (metallogenic zone, i.e. naturally contaminated area) in the dataset that greatly increases the TUF and worsens the sensitivity of the detection of anomalies by a univariate analysis. Boxplots are thus not always suitable for a detailed evaluation and methods such as ECDFs or FDA of PDFs in the framework of Bayes spaces that utilise more information from the data should be preferred. While both ECDF and PDF representations of the concentration distributions are equivalent as such, the clr transformation, which is defined for PDFs, should better capture anomalies within small concentration values. The reason is that the clr transformation is defined  to take into account the relative scale of PDFs, which highlights the fact that the small PDF values are the main source of variability in the concentration density.\\
\begin{figure}[t!]\label{fig:fig_5}
\centering
\includegraphics[width=0.9\textwidth]{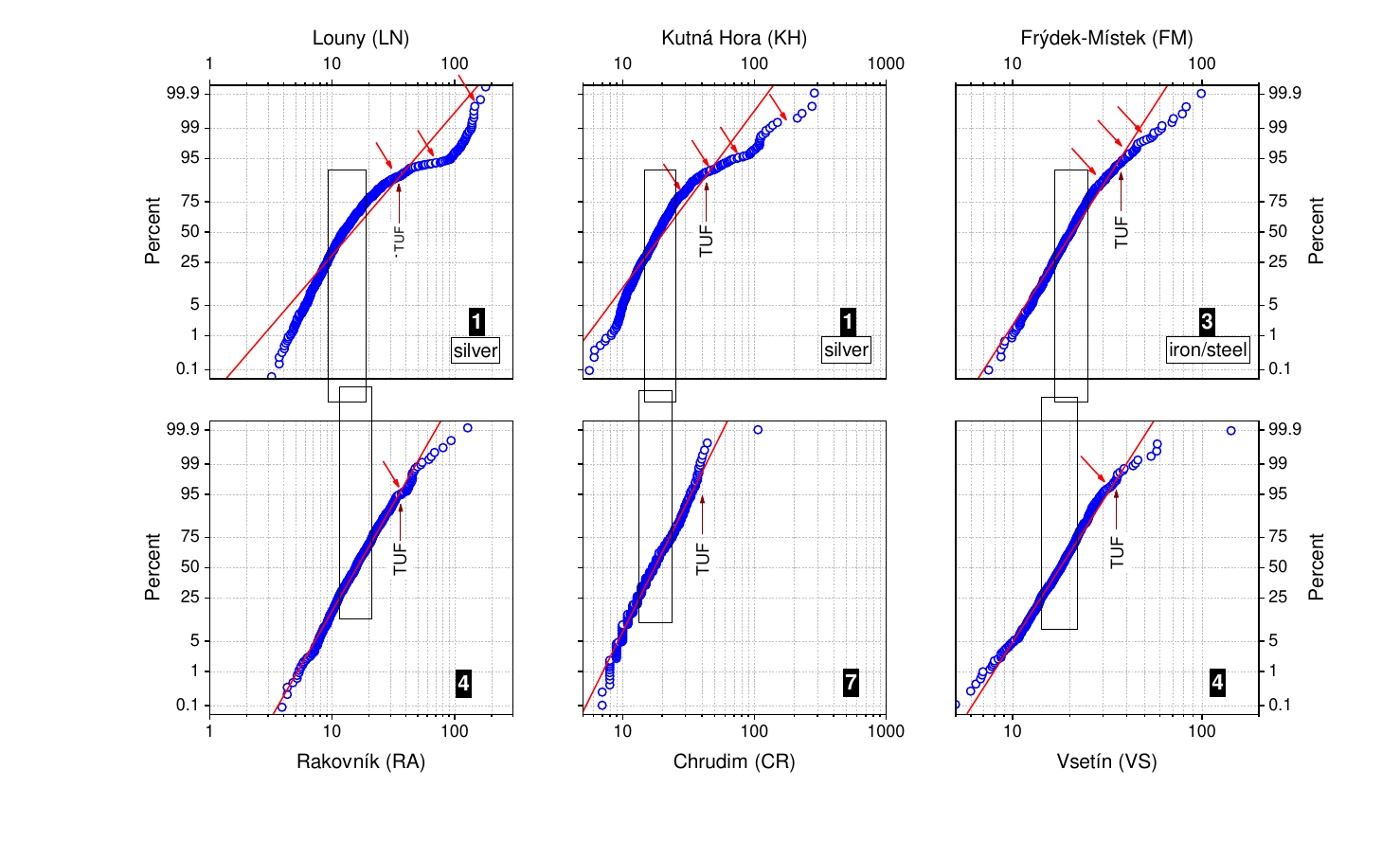}
\caption{ECDFs of Zn concentrations (mg kg$^{-1}$) in selected districts. Captions and explanations of abbreviations are analogous to those in Figure \ref{fig:fig_4}. 
}
\label{fig:fig_5}
\end{figure}

In ECDF stacks (Figures \ref{fig:fig_4} and \ref{fig:fig_5}), selected districts that are well known (or suspected) to be contaminated are plotted in the upper row; the lower row displays plots for neighbouring, less contaminated districts with very similar geology and geography. The vertical pairs of ECDFs are thus suitable for very detailed individual evaluation of concentration patterns. In ECDFs for Cu (Figure \ref{fig:fig_4}), PS and CR districts show monomodal distributions with only exceptional high-value outliers, while the RA district (neighbour of PS) and KH district (neighbour of CR) have much more skewed distributions at high values and considerable occurrences of very high Cu concentrations relative to the main populations of the district, that all resulted in significant departures from the fitted lines at concentrations above the values corresponding to Q3. In KH and UH districts, already the Q3 boundaries are elevated relative to the Q3 in the neighbouring districts that are less contaminated (CR and ZL, respectively), documenting increased concentrations in at least the entire Q4. UH and ZL districts are, however, both anomalous in the entire Czech Republic comparison: their main populations are close to the entire Czech Republic TUF (their bedrock geology includes Cu-rich marine sediments), again underpinning the need to evaluate districts with anomalous background with particular care. \\

Similar features as for Cu can be found in ECDFs for Zn in pairs of selected districts (Figure \ref{fig:fig_5}). ECDFs (Figure \ref{fig:fig_5}) show important diagnostic features of contamination: (i) slower decrease in occurrence of high concentrations in Q4, sometimes starting already in Q3 in considerably contaminated districts, (ii) presence of heterogeneities (minor populations of high concentration values) shown by red arrows in Figures \ref{fig:fig_4} and \ref{fig:fig_5}, and (iii) considerably larger percentage of extreme outliers in the districts contaminated by ore mining and smelting (LN and KH). Those features are not explicitly dependent on the position of the main mode in the concentration distribution that determines the TUF.

\subsection{Univariate data analysis using PDFs and its use for clustering of compartments}\label{sec:anomaly}
\begin{figure}[t!]
\centering
\includegraphics[width=0.9\textwidth]{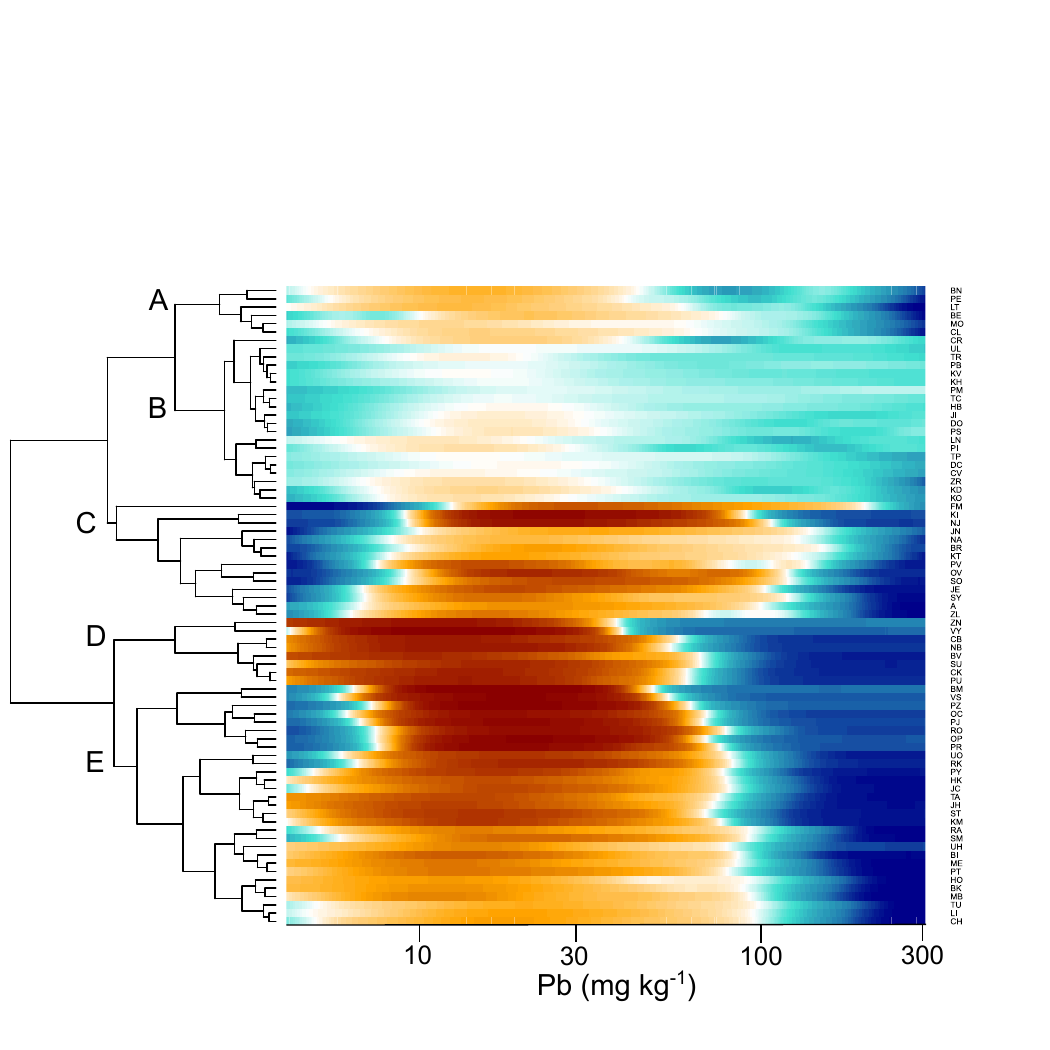}
\caption{PDFs of Pb concentrations, row-wise per district, showing the distribution of concentrations in a colour scale in individual districts and clustered accordingly.}
\label{fig:fig_6}
\end{figure}

The concentration distributions in districts can also be compared using PDF heatmaps of clr transformed arithmetic marginals of univariate concentration series (Figure \ref{fig:fig_6}), i.e. assuming (unlike for geometric marginals) patterns that correspond to one element at a time. Here the densities of concentrations per district are transformed to a colour scale, with the horizontal axis corresponding to element concentrations. Districts with narrow main concentration populations and without contamination are those with warm colours in a narrow range of low concentrations in the heatmaps and sharp colour change to cold hues at the high concentration end. Contamination (i) broadens the main population and/or (ii) leads to heavy tails above the modal maximum with the gradual decline of the data density at the high end.\\

Subsequent clustering of districts according to the PDF representation of their Pb concentrations (Figure \ref{fig:fig_6}) compares more features of concentration distributions than captured by the EDA tools described above. Clustering groups, whose number was determined visually from the dendrogram, recognised districts with higher main modes (Figure \ref{fig:fig_6}, cluster C) together, including JN discussed above (Figure \ref{fig:fig_3}), with the relatively sharp decline of the density above the main mode and thus little contamination, in spite of high overall concentrations. Figure \ref{fig:fig_6} also shows districts with lower main modes and additionally with the sharp decline of densities at concentrations higher than the main mode (sharp change to cold colours), documenting both low background and negligible contamination (Figure \ref{fig:fig_6}, cluster D). Severe contamination is shown by relatively high densities at high concentrations (light blue shades on the right side of the heatmap, Figure \ref{fig:fig_6}, cluster B), actually including districts known to be affected by types 1, 4, and 5 contamination (Table \ref{tab:tab_1}). Noteworthy for that cluster is the broad main mode that is shown by considerably low colour contrast in the entire concentration range. This would make the use of the TUF in these districts inefficient (insufficiently sensitive) due to a large IQR. Particular attention in Figure \ref{fig:fig_6} should be paid to cluster A, which shares similar features with cluster B but with a bit less broadened concentration distribution and weaker contamination, i.e. decay of the concentration densities above the main mode occurring at somewhat lower concentrations than in cluster B. The clusters A and B thus share features expected for areas contaminated by Pb.\\

FDA can also be used for the detection of anomalies in the data set. Here, the DDC algorithm is applied to the arithmetic marginals of Pb data set represented via a matrix of B-spline coefficients. These coefficients are assigned their level of anomaly between -1 (maximal deviation in the negative sense) and 1 (maximal deviation in the positive sense) via scaled residuals computed in step 3 of the algorithm sketched in Section~\ref{sec_ddc}.  

\begin{figure}[h!]
\centering
\includegraphics[width=0.8\textwidth]{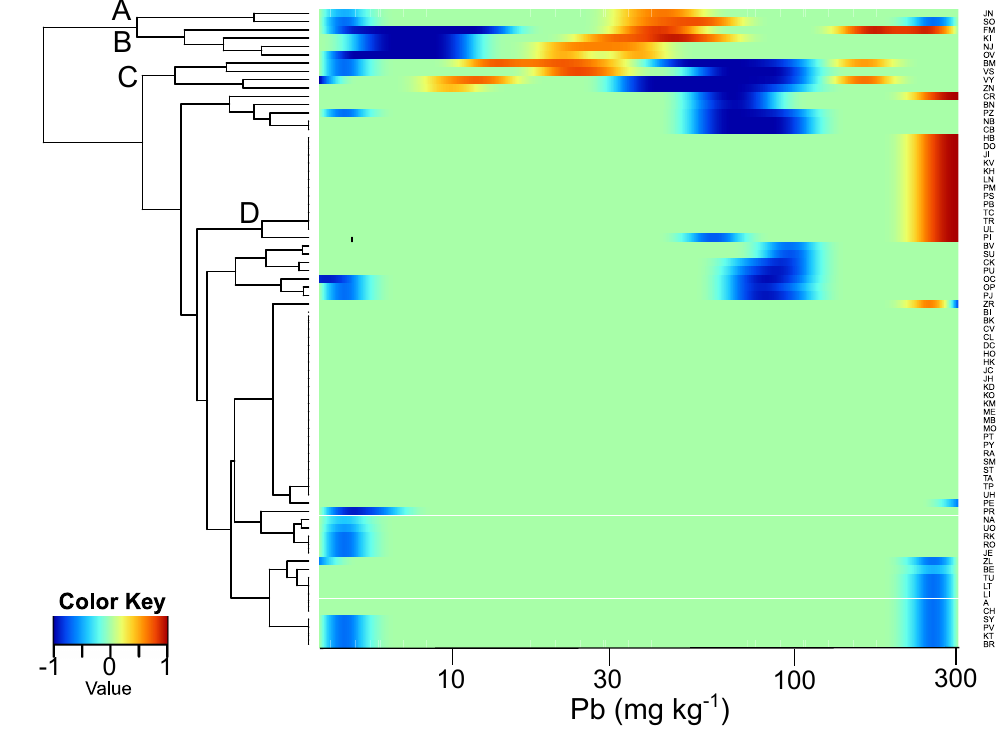}
\caption{Districts clustered according to anomaly detection by the DDC algorithm applied to arithmetic marginals of the Pb distribution.
}
\label{fig:fig_7}
\end{figure}

To visualise the anomaly in the sense of functions rather than coefficients
, the levels of the anomaly were weighted using the functional values of the B-spline basis for a discretised grid of time points. The matrix of weighted anomaly levels then serves as the input for hierarchical clustering, as shown in Figure \ref{fig:fig_7}.

Districts severely impacted by ore mining or industry (contamination types 1, 4, and 5) are identified by the presence of local outliers at concentrations that are one order of magnitude higher than the global background of ca. 20 mg kg$^{-1}$ (cluster D in Figure \ref{fig:fig_7}). The districts JN (and SO) with considerably high background Pb concentration are also indicated as anomalous (cluster A in Figure \ref{fig:fig_7}) although they are not contaminated anthropogenically (type 6 in Table \ref{tab:tab_1}). In clusters B and C, the districts with anomalies at high concentrations (FM, BM, and VY) are really contaminated anthropogenically (types 3 and 4 in Table \ref{tab:tab_1}). Several contaminated samples per district are sufficient to be flagged as anomalous by the DDC algorithm (Figure \ref{fig:fig_7}).

\subsection{Contamination as heterogeneities at high concentrations and a multivariate analysis}
\label{sec:multi}

The traditional EDA definition of anomalies (in particular high anomalies) is based on distinguishing cluster(s) of high concentrations above the main population that are separated from the main population by breaks or gaps in the ECDF \citep{SINCLAIR1991}. Those traditional breaks or gaps are marked by red arrows in the ECDF plots for Cu (top row in Figure \ref{fig:fig_4}) and Zn (Figure \ref{fig:fig_5}). Although placing the arrows into ECDFs is subjective because the actual shape of the natural data distribution for the majority of concentration values is unknown, the presence of some heterogeneity at the high end of the distribution is obviously a diagnostic sign of contamination. Coordinates of the breaks or gaps in ECDFs in individual districts show a remarkable scatter of (i) concentration values at those thresholds and (ii) percentiles of anomalies above those thresholds (portions of anomalously high samples), again underpinning the need for compartmentalisation and an individual approach based on smaller areas. Noteworthy is the variable (district-specific) position of the TUF and of the breaks or gaps in the ECDFs in Figures \ref{fig:fig_4} and \ref{fig:fig_5}, documenting the insufficiency of the TUF to capture the real, fine structure of concentration distributions. It is thus necessary to define local anomalies {within} each particular distribution, represented either using ECDFs or PDFs, without a priori assumptions on the shape of the distribution in the main concentration population or fraction of anomalies.\\

While FDA of PDFs can characterise heterogeneities as those marked by arrows in ECDFs (Figures \ref{fig:fig_4} and \ref{fig:fig_5}) in a different way as shown already in the previous section, it has the additional large advantage of allowing an extension to the multivariate case. Using multivariate PDFs allows the joint distribution of several elements to be characterised simultaneously, which would be hardly possible with traditional EDA tools.\\

\begin{figure}[]
\centering
\includegraphics[width=0.8\textwidth]{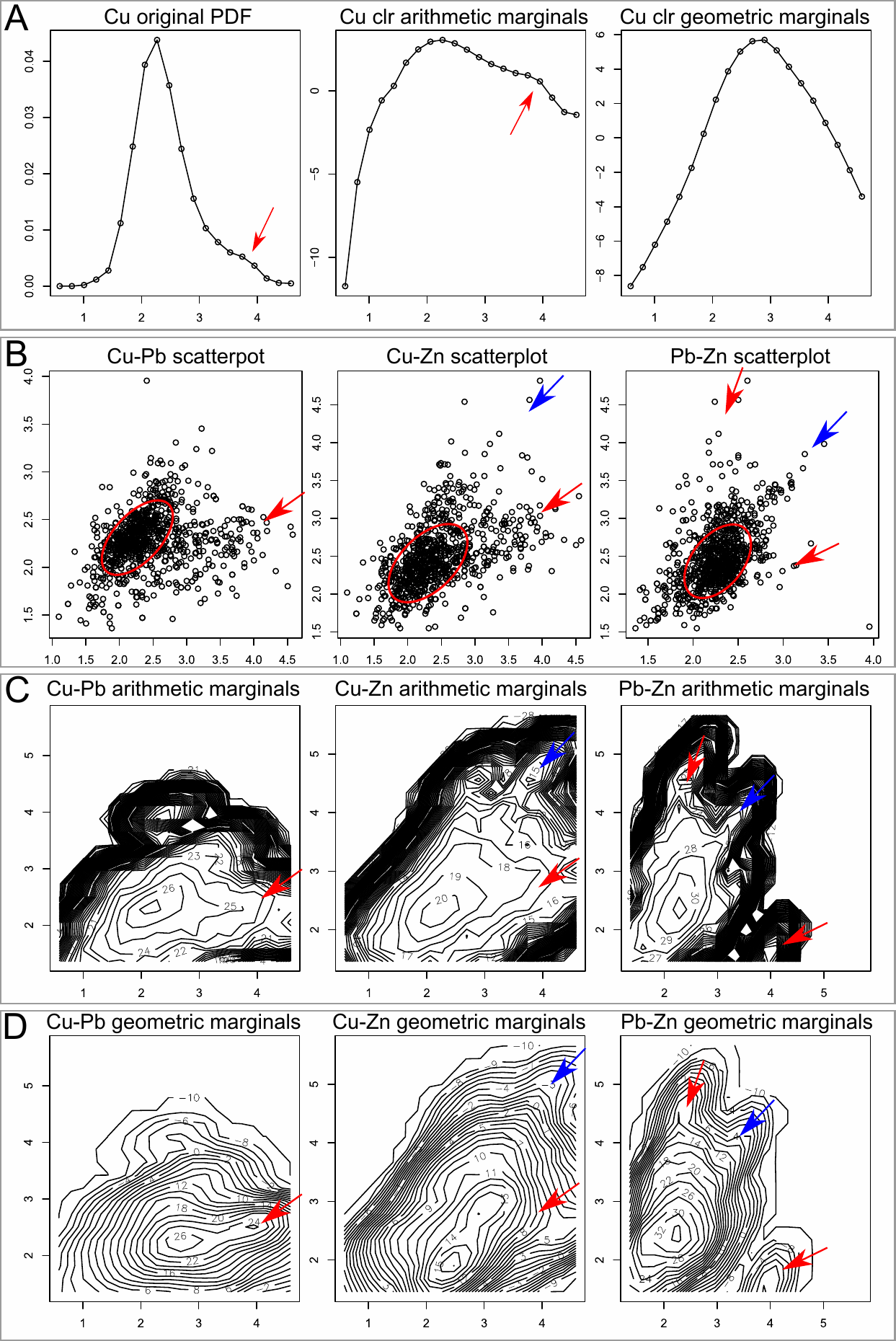}
\caption{Representations of Cu distribution densities (A), scatterplots of pairs of elements (B), and the respective clr transformed bivariate PDFs as arithmetic (C) and geometric marginals (D) for Břeclav (BV) with contamination type 2 (Table \ref{tab:tab_1}). Red arrows show populations of high Cu concentrations unrelated to elevated Pb nor Zn, i.e. single-element contamination. Minor bimetallic contamination by Pb and Zn is highlighted by a blue arrow. Red ellipses in panel B correspond to the robust, MCD covariance ellipses. They highlight the main concentration population for the district and indicate a positive interelement association.}
\label{fig:fig_8}
\end{figure}

\begin{figure}[]
\centering
\includegraphics[width=0.8\textwidth]{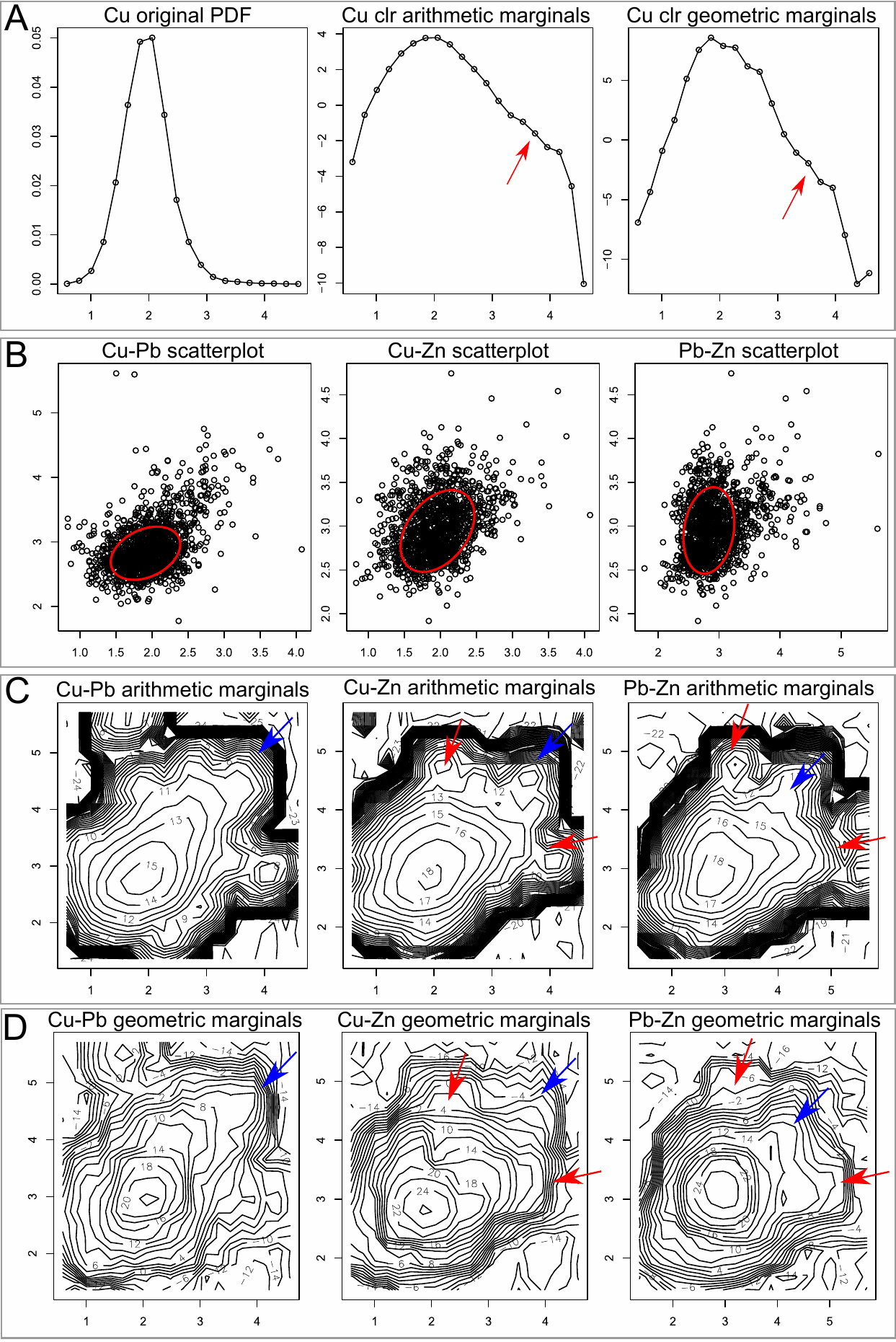}
\caption{Representations of Cu distribution densities (A), scatterplots of element pairs (B), and the respective clr transformed bivariate PDFs as arithmetic (C) and geometric marginals (D) for Jihlava (JI) district with polymetallic contamination type 1 (Table \ref{tab:tab_1}). Blue arrows show samples with multi-element contamination. Red ellipses in panel B correspond to the robust, MCD covariance ellipses. They highlight the main concentration population, and show only weak lithogenic correlation between Cu, Pb, and Zn. 
 }
 \label{fig:fig_9}
\end{figure}

Clr transformed PDFs can find heterogeneities in univariate data distributions, as demonstrated in panels A in Figures \ref{fig:fig_8} and \ref{fig:fig_9} and highlighted there by red or blue arrows. Smoothing of PDFs can be additionally applied to identify bi- and trivariate heterogeneities (panels C and D in Figures \ref{fig:fig_8} and \ref{fig:fig_9}). Figure~\ref{fig:fig_8} shows plots for typical Cu-pesticide impacted districts (BV, type 2 contamination in Table \ref{tab:tab_1}) with a highly heterogeneous distribution of Cu (red arrows in Figure~\ref{fig:fig_8}A). Here the contamination in Cu is truly monometallic (univariate), without the substantial impact of other elements on the Cu concentration at the high end. It is, however, important to take into account some lithogenic effects in the main concentration population controlling jointly all three elements (red ellipses in Figure \ref{fig:fig_8}B). Neither ore mining nor heavy industry have impacted the BV district and neither Pb nor Zn were used in pesticides in vineyards in the Czech Republic. This justifies the use of arithmetic rather than geometric marginals. The heterogeneity of the Cu concentration distribution is amplified in clr transformed arithmetic marginals that better highlight the variability in the outskirts of the data distribution (Figure \ref{fig:fig_8}A, middle panel), similarly as in ECDFs of raw concentration data. Because in BV elevated Cu is not related to high values of Pb and Zn, there are clouds of elevated Cu marked by thick red arrows in the scatter plots (Figure \ref{fig:fig_8}B) outside the main concentration population (ellipses in Figure \ref{fig:fig_8}B). This is reflected by the bivariate PDFs (in terms of clr transformed arithmetic marginals), which highlight these heterogeneities as islands of higher density values around the main population (red arrows in Figure~\ref{fig:fig_8}C and D).\\ 

Visualisation of multimetallic contamination is shown in Figure \ref{fig:fig_9} for the Jihlava district (JI), with mixed contamination from mining polymetallic sulfide ores and silver smelting in the city of Jihlava from the Middle Ages to the 18th century. Here, elevated Cu is associated with elevated Pb and less with elevated Zn (Figure \ref{fig:fig_9}B) and highlighted by multiple maxima in the density maps on the clr-scale (Figure \ref{fig:fig_9}C). While the arithmetic marginals (Figure \ref{fig:fig_9}C) are still reasonable here,  geometric marginals (Figure \ref{fig:fig_9}D), which show similar patterns, better reflect the impact of multimetallic contamination in the univariate and bivariate plotting.

\begin{figure}[]
\vspace{-5.5cm}
\centering
\includegraphics[width=0.9\textwidth]{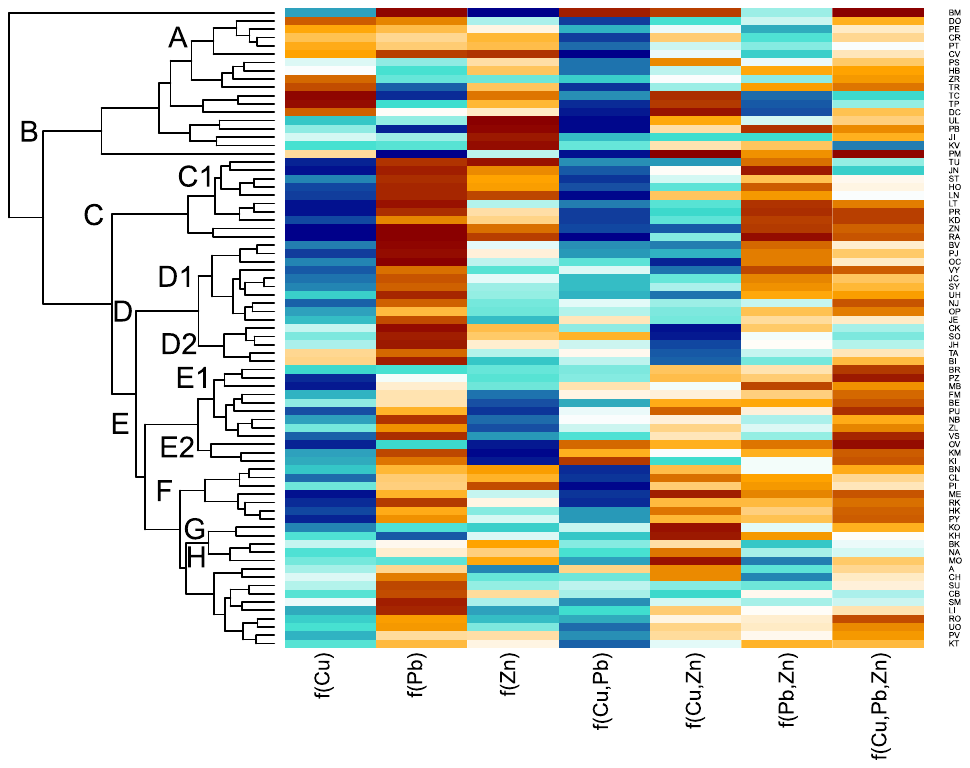}
\vspace{-6cm}
\caption{Compositional clustering of districts according to their respective clr transformed information compositions. Specific values of norms indicate heterogeneity of the respective distributions: low norms (in blue hues) correspond to high heterogeneities, while high norms (in red hues) evidence narrow distributions.}
\label{fig:fig_10}
\end{figure}

The heterogeneity of the densities can be quantified via norms and used for fingerprinting individual districts according to mono-, bi-, and trimetallic distribution heterogeneities. Figure \ref{fig:fig_10} shows districts clustered according to clr transformed information composition vectors defined in \eqref{eq:ic}. Their components corresponding to the univariate geometric marginals are denoted $f(\mathrm{Cu})$, $f(\mathrm{Pb})$, and $f(\mathrm{Zn})$, to the bivariate interactions  $f(\mathrm{Cu},\mathrm{Pb})$, $f(\mathrm{Cu},\mathrm{Zn})$, and $f(\mathrm{Pb},\mathrm{Zn})$, and to the trivariate interaction $f(\mathrm{Cu},\mathrm{Pb},\mathrm{Zn})$. Bi- and trivariate interaction heterogeneities clearly cannot be examined by univariate ECDFs. Yet they are promising indicators of contamination according to specific interelement relationships (Table \ref{tab:tab_1}) and as such extend the EDA diagnostic tools to recognise contamination. For example, districts contaminated by Cu-bearing pesticides (type 2 contamination in Table \ref{tab:tab_1}, districts RA and UH in Figure \ref{fig:fig_4}, and district BV in Figure \ref{fig:fig_8}) are special because of the high heterogeneity of the Cu distribution and simultaneously its mono-metallic character, that in turn produce high bivariate interaction heterogeneities Cu-Pb and Cu-Zn (i.e. these interactions do not indicate a strong uni-, or multimodal pattern and are thus rather uninformative). The bi- and trivariate heterogeneities (blue arrows in Figure \ref{fig:fig_8}) are sensitive to monometallic contamination that pushes points outside the red ellipses in Figure~\ref{fig:fig_8}, panel B, as the ellipses show joint lithogenic controls of element composition. The simultaneous heterogeneities in uni- and bivariate densities with Cu are joint features of clusters C with RA and other pesticide-contaminated districts (LT, PR, and ZN) and D1 with BV, OC, VY, and UH in Figure \ref{fig:fig_10}. In some cases, specific Cu contamination also increases trivariate interaction heterogeneity, such as in the case of districts HO and LN in cluster C1 (Figure \ref{fig:fig_10}).

Clusters B (without subcluster A) and G in Figure \ref{fig:fig_10} include districts with highly heterogeneous Pb distribution, in particular those with ore mining (type 5 in Table \ref{tab:tab_1}) HB, ZR, TC, PB, JI (Figure \ref{fig:fig_9}), and KH (Figure \ref{fig:fig_3}), big cities UL, PM and DC with abundant high Pb pollution, and districts, in which there are two types of soils: normal ones and those in the metallogenic zone of the Ore Mountains -- this is the case for TC (Figure \ref{fig:fig_3}) and KV. Distinguishing anthropogenically contaminated districts and such geologically heterogeneous districts is, however, out of the scope of this exploratory analysis, as both can exhibit the same element concentration patterns. \\

\begin{figure}[h!]
\centering
\includegraphics[width=0.65\textwidth]{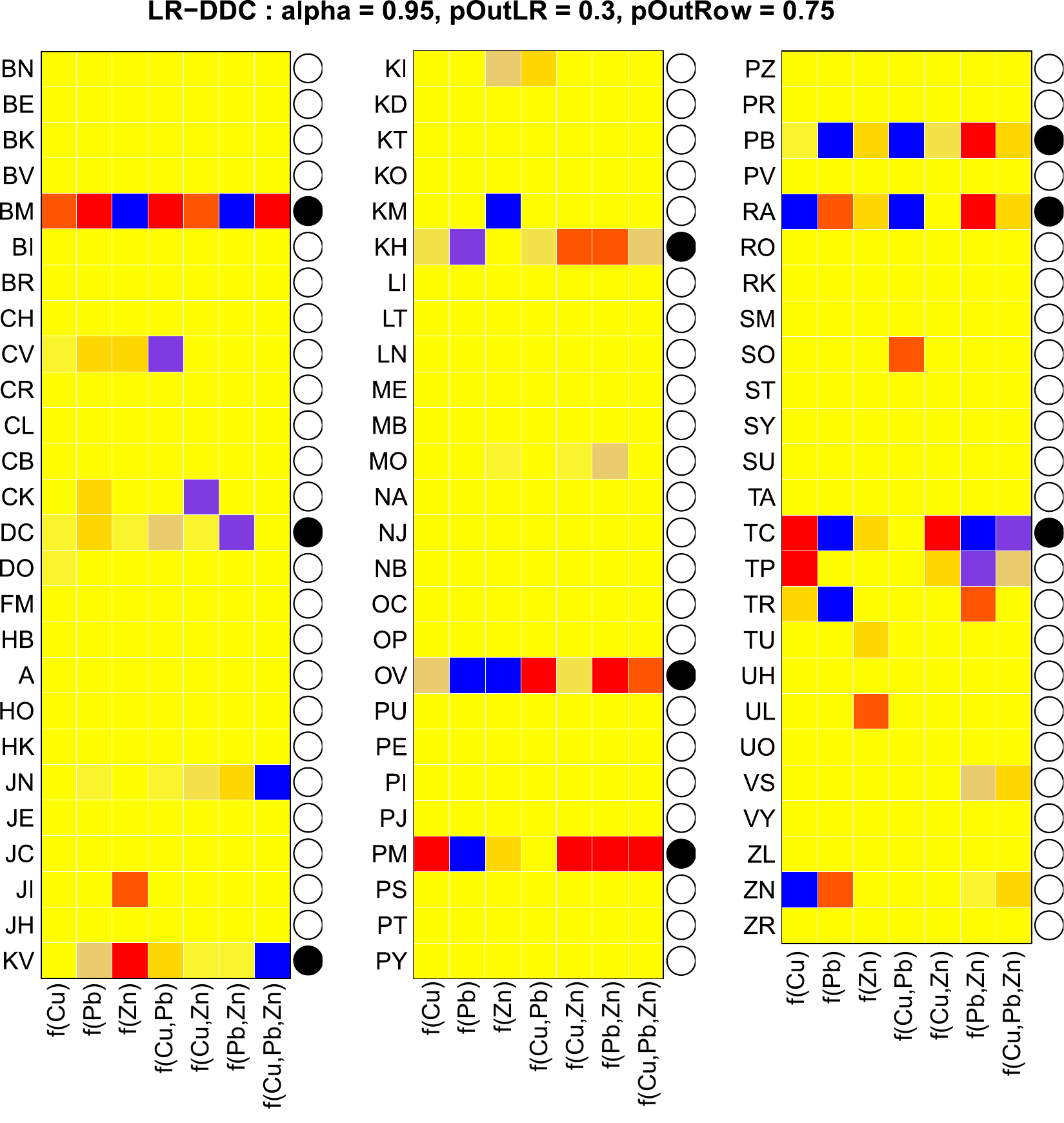}
\caption{Anomaly levels of information composition vectors based
on the scaled residuals of pairwise logratios and aggregated in sense of the original components. The color key corresponds to the case of Figure \ref{fig:fig_7}.
}
\label{fig:fig_new}
\end{figure}

Additionally, the DDC algorithm from Section \ref{sec_ddc} was applied to the information composition vectors as shown in Figure \ref{fig:fig_new}.
Due to compositional nature of the data objects, the original algorithm was adapted to work on all pairwise logratios, while the graphical output is still interpretable in sense of the original components \citep{simicek23}; accordingly, we refer to LR-DDC algorithm.
By following Section \ref{sec:anomaly}, the level of
anomaly is assigned to each pairwise logratio of the information composition, and these levels are aggregated according to each compositional part. The corresponding cell in the graphical output is highlighted if at least some proportion of the corresponding logratios (here 30\%, in line with the default setting) is marked as outlying. Based on the results presented in Figure \ref{fig:fig_new}, each district flagged by LR-DDC as an outlier has contamination with some specific characteristics. BM, OV, and PM are large cities with severe polymetallic contamination (contamination type 4) with particularly high homogeneity in trivariate densities due to all three elements jointly enriched near the contamination sources. The majority of the flagged districts have been impacted by ore mining (KV and PB, contamination type 5) and silver production (KV, KH, and PB, contamination type 1), which left severe contamination but with distinct character. PB is a Czech district with the most severe Pb pollution near the major smelter (high heterogeneity in Pb), with high Pb associated with high Zn (polymetallic contamination, type 5). TC has been contaminated by Pb and Zn from several important metallurgy-related sources of contamination (contamination types 1 and 3), and thus this district shows non-overlapping hotspots of those elements typical of high heterogeneity in polymetallic patterns. RA district is flagged in Figure \ref{fig:fig_new} because of the pesticide-born Cu contamination discussed above.

\subsection{Specificity of diffuse contamination and how to recognise it}

Diffuse contamination (types 2 and 3 in Table \ref{tab:tab_1}) is common in industrialised countries. It refers to a weak, large scale, and thus poorly spatially constrained elevation of risk elements, with continuous increase of risk element concentrations above the natural levels of the main population and no sharp threshold between uncontaminated and contaminated samples. In ECDF, diffuse contamination is visualised as a bend of the main trend in Q4 or even already in Q3 (Figure \ref{fig:fig_4}, district KH; Figure \ref{fig:fig_5}, district FM). Diffuse contamination can also be accompanied by heterogeneities at the high end of the concentration distributions (Figures \ref{fig:fig_4} and \ref{fig:fig_5}, majority of contaminated districts in the upper lines show this feature), but this heterogeneity can generally be lacking or it is poorly discernible (Fig. \ref{fig:fig_4}, district UH) and thus cannot be used as an unequivocal diagnostic sign for diffuse contamination. True diffuse contamination is more resistant to unequivocal identification because it is relatively weak and widespread and thus coalesced with the main concentration population.\\

Diffuse contamination can be identified by expert-based comparison of contaminated districts with neighbouring ones that are uncontaminated and have the same geology and topography. Such pairs are shown in columns in Figure \ref{fig:fig_4}, where UH has been impacted by Cu from pesticide use in vineyards (contamination type 2), while ZL is not suitable for vine production. The diffusively contaminated district UH shows already an elevated Q3 relatively to ZL, while the ECDF of UH has only a poorly discernible change in slope. Here the multi-element pattern helps: Single-element contamination in UH is highlighted by more heterogeneity in the joint density of Cu and Zn relative to ZL; this is why these districts belong to different clusters in Figure \ref{fig:fig_10} (UH and ZL are present in distinct clusters D1 and E, respectively). \\

In contrast to monometallic contamination, the elevated concentrations of two or all three examined elements are associated when there is multi-element contamination, and their bi-element interaction heterogeneities are lower than in the absence of contamination. Diffuse contamination by Pb and Zn jointly (type 3 in Table \ref{tab:tab_1}) impacted the FM district; here indeed the main population has an increased Q3 and more samples above the breaks or gaps in the ECDF of Zn relative to the neighbouring district VS (Figure \ref{fig:fig_5}). An elevated main mode and diffuse enrichment of Pb in FM is shown in Figure \ref{fig:fig_6}, where the districts FM and VS belong to different clusters C and E, respectively. Otherwise, FM and VS districts have similar element patterns and both belong to cluster E in Figure \ref{fig:fig_10}. However, in line with the contamination type, the univariate Pb and Zn distributions are more heterogeneous while the bivariate interaction between Pb and Zn is more homogeneous in FM relative to VS. In any case, Pb concentrations in the FM district are themselves sufficiently high to be evaluated as anomalous by the DDC algorithm (cluster D in Figure \ref{fig:fig_7} with high Pb concentrations). Clearly, all EFDA tools presented above provide complementary information on the univariate and joint variability of the element patterns by considering the main populations (represented by ellipses in panels B, Figures \ref{fig:fig_8} and \ref{fig:fig_9}) as well as the respective anomalies. 

\section{Discussion}
\subsection{Novelty}
Contamination is frequently associated with high outliers for which TUF (or CUF) has become an almost routine tool in EDA \citep{Reimann2018,MatysGrygar2023}. While the TUF approach is satisfactory for the preliminary data examination and swift visualisation in maps, it has considerable limitations. The first is the assumption that the lognormal distribution is the underlying “truth” for the TUF, which seems to be too naive \citep{Reimann2000}. The inadequacy of the TUF for real-life datasets is demonstrated in Figure \ref{fig:fig_3}. The ECDF approach \citep{SINCLAIR1991} is more adequate for real data sets as it (i) preserves the fine structure of the data distributions and (ii) works with them individually (Figures \ref{fig:fig_4} and \ref{fig:fig_5}), irrespective of how large a portion of samples belongs to the main population and how many are anomalous \citep{SINCLAIR1991}. The novelty brought by the FDA of PDFs using the Bayes space approach and demonstrated in the Results section are (i) detailed graphical presentation of the fine structure of univariate data that takes changes in small PDF values into account (Figure \ref{fig:fig_6}) and (ii) clustering of similar patterns in multivariate data structures to simplify the data interpretation (Figure \ref{fig:fig_10}). Figure \ref{fig:fig_6} is an example of how the univariate structure of Pb concentration data can be visualised using PDFs in the clr space while preserving information on the data distribution. The colour scale in Figure \ref{fig:fig_6} highlights the main populations of Pb with a skewed high end and an elevated proportion of high outliers that are not affected by the highly variable main modes in the individual distributions (clusters B and A in Figure \ref{fig:fig_6}). Districts with high natural backgrounds such as JN (Figure \ref{fig:fig_3}) are not considered contaminated by this approach (Figure \ref{fig:fig_6}), although their main mode is anomalous as compared to other Czech districts (Figures \ref{fig:fig_3} and \ref{fig:fig_7}).\\

Another novelty the FDA of PDFs brings to environmental geochemistry exploration is a visualisation and quantification of multivariate heterogeneities in data distributions without making explicit assumptions on what constitutes the “ideal” data distribution. The heterogeneity of the data distribution, whose identification is a key feature of the ECDF analysis, is captured by compositional data analysis of the information compositions. Such analysis can help to identify monometallic diffuse contamination by Cu pesticides because it increases both heterogeneities of the Cu distribution and of the bivariate distribution of Cu and Zn (Figure \ref{fig:fig_10}). Identification of diffuse contamination is a challenge for soil data exploration as it needs a comparison of contaminated and uncontaminated areas with similar factors controlling the soil composition (Figures \ref{fig:fig_4} and \ref{fig:fig_5}). In contrast, Cu contamination from ore mining and smelting is usually polymetallic and thus it is characterised by lower heterogeneity of the joint distributions of Cu, Pb, and Cu, Zn, respectively.\\

The methodology introduced in this paper could be applied to multivariate analyses including more than three elements. That extension would be technically feasible, but the geochemical interpretation of results could be more challenging. The choice of Cu, Pb, and Zn in this paper was not arbitrary: those elements have similar fates in soils (grain-size control, mobility, adsorption to particles) and thus similar natural controls, which makes their relationships rather simple and understandable. Further risk elements, that could be attractive in contamination tracing, show many specific features, such as larger mobility in soil profiles (Cd), strong control by local geology (As), and prevalent spread by the atmosphere (As, Hg). Those specificities would weaken interelement relationships relative to the Cu, Pb, and Zn triad. Still, multivariate EFDA would deserve testing, for example for the identification of polymetallic diffuse contamination.

\subsection{From big datasets to compartments and back to the whole}

The majority of environmental geochemistry studies apply the same thresholds to the entire available dataset \citep{Toth2016, Reimann2018} or evaluates ECDFs for entire datasets \citep{FabianKarl2017}. Such a “mass processing” of big data must inevitably destroy details of the entire picture. \citet{Negrel2015} and \citet{MatysGrygar2023} concluded that the interpretation of big geochemical datasets is challenging due to too broad scale of factors controlling the soil composition. Our idea for the Czech Republic data was thus to separate the big dataset and “zoom in” on smaller areas, assuming that factors that control the soil composition are more homogeneous and thus more easily deciphered \citep{MatysGrygar2023}. The idea of zooming in is generally opposite to the prevailing current trend of processing big data by brute computational force of explanatory models, frequently working in a black box manner. The risks of this general trend have been critically discussed by \citet{Rudin2019} and \citet{Loyola-González2019}. Together with \citet{Rudin2019} we promote the use of interpretable models, also called white-box models \citep{Loyola-González2019}. \\

In this paper, the big soil chemistry dataset of the entire Czech Republic was divided into compartments, which are historically established administrative districts, typically with certain natural geographic boundaries, a rather homogeneous geology, and consistent agricultural activities. \citet{Ballabio2018} used NUTS2 regions for compartmentalisation of their pan-European dataset, which was a good step but perhaps a bit too coarse: In \citet{Ballabio2018} the entire Czech Republic was divided into eight NUTS2 regions, boundaries of which were defined by Eurostat, while in this work, we used 76 districts, i.e., we worked on a much finer spatial scale. The compartmentalisation of big datasets would considerably extend the volume of work needed for data processing if it needs to be done manually using traditional EDA tools \citep{SINCLAIR1991}. If this is not done, only several contaminated areas can be identified in the whole Czech Republic as was demonstrated on a low-resolution map by \citet{Bednarova2016}. Novel FDA tools from \citet{Genest2023}, tailored for the use in geochemistry in this paper, in particular their powerful visual representations and clustering according to internal similarity, make it possible to zoom in without losing the fine structure of the data set. The power of the Bayes space approach is documented by a graphical representation of Pb concentration distributions for various districts in the Czech Republic (Figure \ref{fig:fig_6}) and district clustering according to uni- and multivariate density heterogeneity (Figure \ref{fig:fig_10}). Of course, the potential of FDA of PDFs goes beyond this initial study, which can be further expanded in many directions \citep{kokoszka17}.

\subsection{Sensitivity test of EFDA for diffuse contamination by Cu pesticides}

The need for larger sensitivity in distinguishing contamination from a highly variable, geographically specific natural background is documented in Figure \ref{fig:fig_3}: the use of the TUF for the entire Czech Republic makes no sense on the district level. Too broad spatial scales and too high thresholds prevented the identification of soil contamination in the Czech Republic by Cu pesticides in vineyards and hop gardens in the maps of \citet{Bednarova2016} and \cite{Ballabio2018}, although it can be visualised if proper data mining tools are used \citep{MatysGrygar2023}. \\

\begin{figure}[]
\centering
\includegraphics[width=0.8\textwidth]{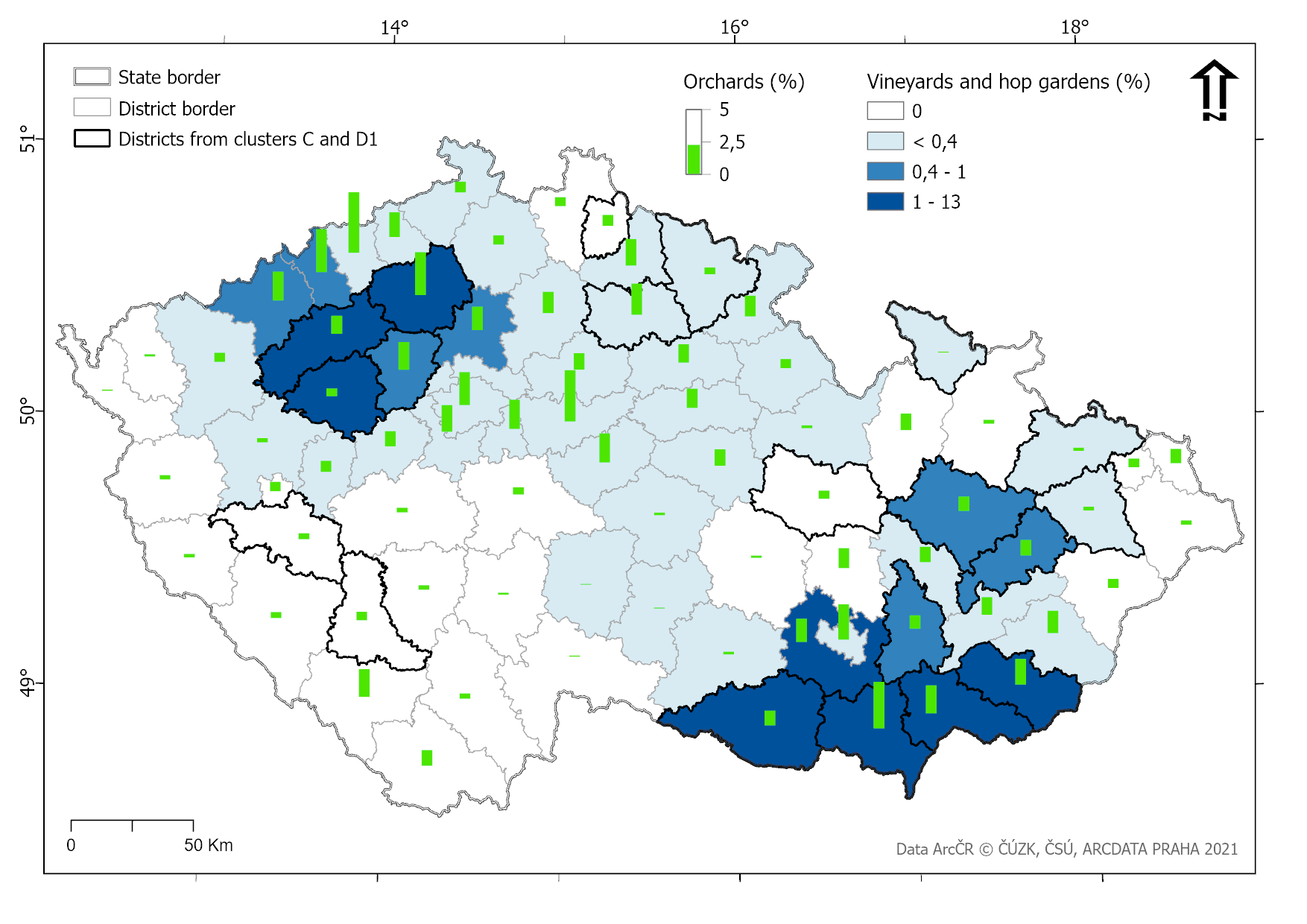}
\caption{Clusters C and D1 (districts bordered by thicker lines), the proportion of hop gardens and vineyards (blue shades) as well as orchards (green columns) in Czech administrative districts.}
\label{fig:fig_11}
\end{figure}

Clusters C and D1 in Figure \ref{fig:fig_10} include districts with high variability in Cu concentrations combined with a relatively weak relation between Cu and Zn; taken together these are diagnostic signs of a monometallic Cu contamination. Those clusters contain 20 districts, in particular all 8 districts with hop gardens and vineyards that actually represent more than 1\% of agricultural soils in those districts. The fraction of hop gardens and vineyards is more than 0.4\% in 15 districts, 12 of which are included in the C and D1 clusters. Figure \ref{fig:fig_11} shows the coincidence of the intensity of hop and wine production and the position of districts with monometallic Cu contamination according to Figure \ref{fig:fig_10}. Possibly also abundance of orchards is relevant, as some fruit trees or shrubs are also treated by pesticides. Part of the districts where the Cu contamination is not related to those specific crops was impacted by metal mining; natural factors may also play a role, such as local mafic rocks \citep{MatysGrygar2023}. Also, information on hop and wine production does not include historical activities. Still, the match of EFDA results and wine- and hop production intensity shown in Figure \ref{fig:fig_11} documents good EFDA performance.

\section{Conclusion}

This work paves new ways to data mining in soil geochemistry datasets. A considerable volume of sampling and analytical work has been performed in the recent decade. However, there is a pressing need to develop powerful data processing tools that would allow us to fully use these data to gain an understanding of natural and anthropogenic element patterns in soils. Regarding the geochemical principles of our approach, we returned to traditional exploratory data analysis, developed for ore prospection by the end of the 20th century, and extended it to a multidimensional analysis. To make big datasets processable, novel tools from the domain of functional data analysis of probability density functions using the Bayes space methodology were developed here. The graphical presentation of results was tailored to process big data without the loss of details and while respecting a wide variety of natural factors that govern the soil composition. The proposed tools, tested on a soil geochemistry dataset of the Czech Republic, can be used in any soil mapping project and can identify anthropogenic contamination with larger sensitivity than conventional data mining tools. For better accessibility of the new concepts, trivariate PDFs were analysed in this paper. The presented mathematical tools could naturally also deal with higher dimensional data, but this would require resolving issues related to data processing (e.g., kernel density estimation becomes computationally intensive in higher dimensions) and interpretation (the complex data structure can have a substantial impact on both geometric and arithmetic marginals and a geochemical interpretation may be elusive). Nevertheless we hope to address these challenges in the near future.

\section*{Acknowledgement}
The authors are grateful to the Ministry of Agriculture of the Czech Republic for providing the soil database RKP for free for academic purposes. J. Elznicov\'a (J. E. Purkyn\v e University in \'Ustí nad Labem, Czech Republic) kindly helped with the geographic data search and processing. T. Matys Grygar, K. Hron, I. Pavl\r u, J.G. Ne\v slehov\'a and U. Radoji\v ci\'c were supported by the Czech Science Foundation, Project 22-15684L, and by the Austrian Science Foundation, Project I 5799-N, respectively. S. Greven and J.G. Ne\v slehov\'a acknowledge funding from the German Research Foundation (DFG grant GR 3793/8-1) and the Natural Sciences and Engineering Research Council of Canada (RGPIN-2022-03614), respectively. Part of this work was completed while K. Hron visited the Centre de recherches math\'ematiques (CRM) in Montr\'eal (Qu\'ebec, Canada) in May 2023 as part of the CRM–Simons Scholar-in-Residence program.\\

Role of the authors: T. Matys Grygar designed the geochemical part of the work and prepared the initial manuscript draft. U. Radoji\v ci\'c performed the preliminary statistical analysis, as well as the first draft of the methodology section, and helped to finalise the manuscript. I. Pavl\r u worked on the outlier detection and prepared the initial draft of the DDC section. S. Greven and J.G. Ne\v slehov\'a worked on the methodological part of the manuscript and helped to finalise the paper. 
\v S. T\r umov\'a performed work with geographic information systems and produced maps. K. Hron worked on the methodological part of the paper, helped to finalise the manuscript, and coordinated the entire teamwork.

\bibliographystyle{MG}       
{\footnotesize
\bibliography{bibliography}}   
\end{document}